\documentclass[12pt]{article}
\pdfoutput=1

\usepackage{cite}

\usepackage{tikz}
\usetikzlibrary{arrows,shapes}
\usetikzlibrary{trees}
\usetikzlibrary{matrix,arrows} 				% For commutative diagram
\usetikzlibrary{positioning}				% For "above of=" commands
\usetikzlibrary{calc,through}				% For coordinates
\usetikzlibrary{decorations.pathreplacing}  % For curly braces
\usepackage{pgffor}							% For repeating patterns

\usetikzlibrary{decorations.pathmorphing}	% For Feynman Diagrams
\usetikzlibrary{decorations.markings}
\usetikzlibrary{snakes}

\tikzset{
    vector/.style={decorate, decoration={snake}, draw},
	provector/.style={decorate, decoration={snake,amplitude=2.5pt}, draw},
	antivector/.style={decorate, decoration={snake,amplitude=-2.5pt}, draw},
    fermion/.style={draw, postaction={decorate},
        decoration={markings,mark=at position .55 with {\arrow[draw]{>}}}},
    fermionbar/.style={draw, postaction={decorate},
        decoration={markings,mark=at position .55 with {\arrow[draw=black]{<}}}},
    fermionnoarrow/.style={draw},
    gluon/.style={decorate, draw,decoration={coil,amplitude=4pt, segment length=6pt}, line width=1},
    scalar/.style={dashed,draw, postaction={decorate},
        decoration={markings,mark=at position .55 with {\arrow[draw]{>}}}},
    scalarbar/.style={dashed,draw, postaction={decorate},
        decoration={markings,mark=at position .55 with {\arrow[draw]{<}}}},
    scalarnoarrow/.style={dash pattern = on 6 pt off 3 pt,draw},
    electron/.style={draw, postaction={decorate},
        decoration={markings,mark=at position .55 with {\arrow[draw]{>}}}},
	bigvector/.style={decorate, decoration={snake,amplitude=4pt}, draw},
	vectorscalar/.style={loosely dotted,draw, postaction={decorate}},
}

\usepackage{relsize}

\usepackage[nointlimits,reqno]{amsmath}
\usepackage{amssymb}
\numberwithin{equation}{section} % Equation numbering within sections
\usepackage{color}

\usepackage{graphicx}

\usepackage[pdftex,
 pdfproducer=TeXShop,
 pdfcreator=pdflatex]{hyperref}

\usepackage{xspace}
\usepackage{luty}

\newcommand{\bea}{\begin{eqnarray}}
\newcommand{\eea}{\end{eqnarray}}

\newcommand{\lag}{\mathcal L}

\begin{document}

% =============================================================================
% Title page
% =============================================================================
\begin{titlepage}
%\preprint{}
\begin{flushright}
  SLAC-PUB-15591
\end{flushright}

\title{Induced Electroweak Symmetry Breaking\\
and Supersymmetric Naturalness}
\author{Jamison Galloway$^a$,\ \ 
Markus Luty$^b$,\ \ 
Yuhsin Tsai$^b$,\ \ 
Yue Zhao$^c$} %,\ \ 

\address{$^a$Dipartimento di Fisica, Universit\`a di Roma ``La Sapienza"\\
{\rm and} INFN Sezione di Roma, I-00185 Roma}

\address{$^b$Physics Department, University of California Davis\\
%One Shields Avenue\\
Davis, California 95616}

\address{$^c$SLAC, Stanford University\\
%One Shields Avenue\\
 Menlo Park, California 94025}

\begin{abstract}
In this paper we study a new class of supersymmetric models
that can explain a $125\GeV$ Higgs without fine-tuning.
These models contain additional `auxiliary Higgs' fields with
large tree-level quartic interaction terms but no Yukawa couplings.
These have electroweak-breaking vacuum expectation values, and
contribute to the VEVs of the MSSM Higgs fields either through an induced
quartic or through an induced tadpole.
The quartic interactions for the auxiliary Higgs fields can arise  from
either $D$-terms or $F$-terms.
The tadpole mechanism has been previously studied in strongly-coupled
models with large $D$-terms, referred to as `superconformal technicolor.'
The perturbative models studied here 
preserve  gauge coupling unification in the simplest possible way,
namely that all new fields are in complete $SU(5)$ multiplets.
The models are consistent with the observed properties of the 125~GeV 
Higgs-like boson as well as precision electroweak constraints, and predict
a rich phenomenology of new Higgs states at the weak scale.
The tuning is less than $10\%$ in almost all of the phenomenologically
allowed parameter space.
If electroweak symmetry is broken by an induced tadpole, the 
cubic and quartic Higgs self-couplings are significantly smaller than 
in the standard model.
\end{abstract}

\end{titlepage}

% =============================================================================
% =============================================================================
\section{Introduction}
\eql{sec:intro}
% =============================================================================
% =============================================================================
The discovery of a Higgs-like particle with mass near $125\GeV$ at the
LHC represents a major advance in our understanding
of electroweak symmetry breaking \cite{Chatrchyan:2012ufa,Aad:2012tfa}.  %  (EWSB).
The couplings of this state to the $W$ and $Z$ are close to that of a
standard model (SM) Higgs, providing direct evidence that this state is
the dominant excitation of the condensate that breaks electroweak symmetry.
Even though the couplings of this state are compatible with those
of a SM Higgs, there 
is still room for significant mixing with other Higgs
states and/or compositeness of the Higgs at higher scales \cite{ATLAS-CONF-2013-034,CMS-PAS-HIG-12-045}.

The $125\GeV$ Higgs-like particle is a mixed blessing for
supersymmetry.
A light Higgs is a hallmark of supersymmetry, but supersymmetric
models generally predict a Higgs \emph{lighter} than $125\GeV$.
In the MSSM, SUSY
relates the maximal tree-level Higgs quartic coupling 
 to the electroweak gauge couplings via
$\la = \sfrac 18 (g^2+ g'^2)$.
This leads to the tree-level Higgs mass bound $m_h < m_Z$, which was already 
% essentially
ruled out by LEP.
Explanations for the observed mass of the Higgs in SUSY have focussed
on additional contributions to the Higgs quartic:
\begin{itemize}
\item {\it MSSM:}
Top/stop loops can generate a large quartic \cite{Ellis:1991zd,Haber:1990aw,Okada:1990vk,Brignole:1992uf,Carena:1995bx,Birkedal:2004zx}.
However, the same loops also generate a large Higgs quadratic term, 
resulting in tuning at the level of at least $1\%$.
\item {\it NMSSM:}
The superpotential coupling $\la S H_u H_d$ gives an additional
contribution to the Higgs quartic that can alleviate the naturalness
problem \cite{Nilles:1982dy,Ellis:1988er,Espinosa:1992hp,Kane:1992kq}.
Taking $\la$ as large as possible consistent with perturbativity below
the GUT scale improves naturalness relative to the MSSM, but improved
naturalness is obtained for larger $\la$ \cite{Cao:2008un,Cavicchia:2007dp,Barbieri:2010pd,Lodone:2010kt,Franceschini:2010qz}.
See \cite{Cao:2012fz,Hall:2011aa} 
for discussions in light of the Higgs discovery.
\item {\it Non-decoupling $D$-terms:}
New gauge interactions broken at the weak scale can give
additional contributions to the Higgs quartic without tuning \cite{Batra:2003nj,Maloney:2004rc}.
These require significant additional matter content to maintain
gauge coupling unification.
\item
{\it Fat Higgs:}
Compositeness of the NMSSM Higgs fields above the weak scale can explain why
couplings like the $\la$ coupling of the NMSSM are large at the weak scale
without Landau poles below the GUT scale \cite{Harnik:2003rs,Chang:2004db}.
These models also require significant additional matter content to
maintain gauge coupling unification.
\end{itemize}
Overall, there seems to be a trade-off between naturalness and simplicity,
leading a number of authors to investigate the possibility that
SUSY is not natural \cite{Wells:2003tf,Giudice:2004tc,ArkaniHamed:2004fb,Giudice:2011cg,ArkaniHamed:2012gw,Arvanitaki:2012ps}.

\begin{figure}[tbp]
\begin{center}
\includegraphics{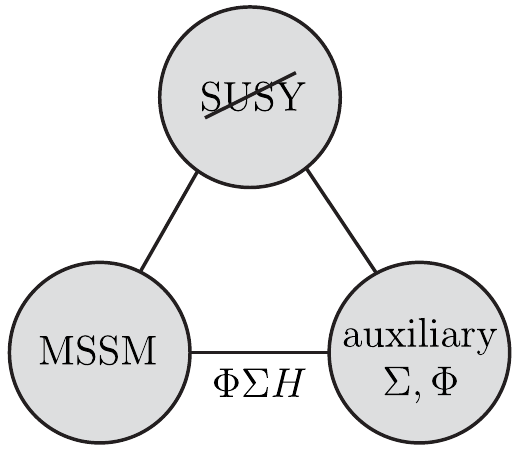}
\caption[]{Schematic structure of the models.
The auxiliary Higgs sector contains electroweak doublets $\Si$ and
electroweak singlets $\Phi$.
These interact with the Higgs fields $H$ of the MSSM via
superpotential couplings and $A$ terms of the form $\Phi \Si H$.}
\label{fig:3site}
\end{center}
\end{figure}

In this paper, we consider a different approach to Higgs naturalness, 
illustrated schematically
in Fig.~\ref{fig:3site}.
%
%%%%%%%%%%%%%%%%%%%
%%%%%%%%%%%%%%%%%%%
%
The idea is that there is an additional sector containing `auxiliary
Higgs' fields with large quartic self-interactions, but no Yukawa
couplings.
These couple to the MSSM Higgs fields via superpotential couplings
and soft SUSY breaking terms, and therefore contribute to the Higgs
potential.
The auxiliary Higgs fields have VEVs that break electroweak symmetry,
so the observed light Higgs is a mixture of the MSSM Higgs and auxiliary
Higgs fields.
In different limits, the dominant effect on the light Higgs can be viewed
as an induced quartic interaction or an induced tadpole.
The latter mechanism was proposed in \Ref{Azatov:2011ps,Azatov:2011ht},
which considered models where the auxiliary Higgs
fields were composites arising from a strong superconformal sector.
The mechanism was therefore called
`superconformal technicolor' (see also related work in \cite{Craig:2011ev,Gherghetta:2011na,Heckman:2011bb}).
In this paper we construct perturbative models of this
mechanism, which we call `induced EWSB'.%

The large quartic interactions for the
auxiliary Higgs fields can arise from new gauge interactions
($D$-terms) and/or new superpotential interactions ($F$-terms).
In the case of $D$-terms, this requires that the auxiliary
Higgs fields be charged under a new gauge group that is
broken at the TeV scale.
The new gauge couplings can easily be stronger than the 
electroweak gauge couplings at the TeV scale,
so the tree-level auxiliary Higgs quartic can be significantly
larger than the tree-level Higgs quartic in the MSSM.
In the $F$-term models, the auxiliary Higgs quartic
arises from a superpotential coupling $\la S \Si_u \Si_d$
between a singlet $S$ and auxiliary Higgs doublets $\Si_{u,d}$.
This can be somewhat larger than the analogous coupling
in the NMSSM because $\Si_{u,d}$ do not
have Yukawa couplings.

Precision gauge coupling unification can be incorporated in
these models in a very simple way:
all fields beyond the MSSM can come in complete $SU(5)$ multiplets.
For the $D$-term models, VEVs of the auxiliary Higgs fields
give rise to mixing between the gauge bosons of the new gauge
interactions and those of the electroweak group.
This is naturally small if the breaking scale of the new gauge
group is sufficiently large, which is in any case required
by precision electroweak constraints. 

We now describe how the auxiliary Higgs fields improve
naturalness of electroweak symmetry breaking.
There are two different limits that can be simply understood
by integrating out a heavy Higgs multiplet.
In the `induced quartic' limit, the dominant effect is a contribution to 
the quartic of the light Higgs.
In the `induced tadpole' limit, it is a tadpole for the light Higgs.

We can exhibit these limits in a simplified
model with two Higgs doublets $\Si$ and $H$,
where $\Si$ is the auxiliary Higgs with
a large quartic and $H$ is the MSSM Higgs field.
The potential is
\begin{align}
\eql{Vsupersimp}
V = m_H^2 |H|^2 + m_{\Si}^2 |\Si|^2 - \ka^2 (\Si^\dagger H + \hc)
+ \la_\Si |\Si|^4.
\end{align}

We first consider the decoupling limit where one linear combination
of $H$ and $\Si$ has a large positive mass-squared.
In this limit the light mass eigenstate is
\begin{align}
H_1 = \sin \ga \, H + \cos \ga \, \Si,
\end{align}
where $\ga$ is the mixing angle that diagonalizes the quadratic terms.
The effective potential is then
\begin{align}
V_{\rm eff} = m_1^2 |H_1|^2 + \la_\Si \cos^4 \! \ga \, |H_1|^4,
\end{align}
where $m_1^2$ is the light mass-squared eigenvalue.
In this limit, the potential has induced a quartic interaction for the
light Higgs.
In the case where $m_2^2 \gg m_1^2$, this limit generally requires a tuning
proportional to $m_1^2 / m_2^2$, but we do not need such an extreme hierarchy 
in a realistic theory.

The other limit we are interested in occurs when $\la_\Si$ is
large and $\ka^2$ is treated as a perturbation.
We focus on the CP-even Higgs bosons and write
\begin{align}
\Si = \frac{1}{\sqrt{2}} \begin{pmatrix}
0 \\ \si
\end{pmatrix},
\qquad
H = \frac{1}{\sqrt{2}} \begin{pmatrix}
0 \\ h
\end{pmatrix}.
\end{align}
For $\ka^2 = 0$, $\Si$ and $H$ decouple, and we have
\begin{align}
\avg\si = f,
\qquad
f^2 = -\frac{m_\Si^2}{\la_\Si}.
\end{align}
The heavy mass eigenstate has mass $m_\si^2 = 2\la_\Si f^2$.
If this mass is sufficiently heavy, 
we can integrate it out to get an effective Higgs potential
\begin{align}
V_{\rm eff} = \sfrac 12 m_H^2 h^2 - \ka^2 f h + O(\ka^4).
\end{align}
Note that a tadpole for $h$ has been generated.
If $m_{H}^2 > 0$, this gives a stable VEV for $h$,
\begin{align}\eql{HVEVtoy}
v_h = \avg{h} = \frac{\ka^2 f}{m_{H}^2}.
\end{align}
We see that electroweak symmetry breaking is dominated by
a tadpole induced by the heavy Higgs.

Let us understand the approximation in the induced tadpole limit
more systematically.
Including higher order terms, the potential obtained by
integrating out $\Si$ has the form
\begin{align}\eql{Veffsupersimptadpole}
V_{\rm eff} \sim m_H^2 h^2
- \ka^2 f h \left[ 1 + \frac{\ka^2}{\la_\Si f^3} h
+ \left( \frac{\ka^2}{\la_\Si f^3} h \right)^2
+ \cdots \right].
\end{align}
From this we see that the higher-order terms can be neglected
if $\ep \ll 1$, where 
\begin{align}\eql{epsuilondefn}
\ep = \frac{\ka^2 v_h}{\la_\Si f^3} \sim \frac{m_h^2}{m_\si^2} \frac{v_h^2}{f^2}.
\end{align}
In the last step we used $m_h^2 \sim m_H^2$, $m_\si^2 \sim \la_\Si f^2$,
and \Eq{HVEVtoy}.
We see that for $f \sim v_h$ the expansion is valid when $m_\si \gg m_h$.
Since $m_\si^2 \sim \la_\Si f^2$, this requires a relatively large
quartic interaction for $\Si$.

A tadpole interaction violates electroweak gauge
symmetry explicitly, and might thus be cause for some suspicion.
However, the point is that such an interaction is allowed because
the VEV of the heavy Higgs breaks electroweak gauge symmetry.
In the low energy effective theory, electroweak gauge symmetry is
nonlinearly realized by Nambu-Goldstone bosons.
For small $\ka^2$, the Nambu-Goldstone fields are contained in $\Si$:
\begin{align}
\Si = \frac{1}{\sqrt{2}} e^{i\Pi / f} 
\begin{pmatrix}
0 \\ f \end{pmatrix}
+ \cdots,
\end{align}
where $\Pi = \frac 12 \tau_a \Pi_a(x) $.
This transforms correctly under electroweak gauge transformations
provided that $\Pi$ transforms nonlinearly in the standard way.
The fields $\Pi$ are light, and the 
effective potential for $H$ can then be written as
\begin{align}\eql{VHsimpnonlinear}
V_{\rm eff} = m_H^2 |H|^2 - \frac{\ka^2 f}{\sqrt{2}} 
\left[ \begin{pmatrix}
0 \\ f \end{pmatrix}^T \!\!
e^{-i\Pi / f} H + \hc \right] + \cdots.
\end{align}
The nonlinear transformation of the 
$\Pi$ fields ensures that the tadpole term is formally invariant under
electroweak gauge symmetry. 
The $\Pi$-dependent terms in \Eq{VHsimpnonlinear} contain
important mixing terms between the CP-odd fields,
but they do not change the results above for the
CP-even fields.
It is straightforward to check that the results
above reproduce the results of minimizing the full potential.

Even away from the limits discussed above, the VEV of $H$ 
in our simplified model can
be thought of as being `induced' by $\Si$, in the following sense.
The mass matrix of the CP-even fields $\si$ and $h$ is
\begin{align}
\eql{masssupersimp}
\scr{M}^2 = \begin{pmatrix}
2\la_\Si f^2 + m_H^2 v_h^2 / f^2 & -m_H^2 v_h/f \\
-m_H^2 v_h/f & m_H^2
\end{pmatrix}.
\end{align}
where we have eliminated $m_\Si^2$ and $\ka^2$ in favor of
$v_h$ and $f$ using the potential minimization conditions.
We have $\det(\scr{M}^2) = 2\la_\Si f^2 m_H^2$, so vacuum stability
{\it requires} $m_H^2 > 0$.
Thus, electroweak symmetry would be unbroken if the
$\Si$ fields were not present.

In the body of the paper we will consider the phenomenology of
induced EWSB in detail.
We will find that the simplified model presented above gives a good
description of the underlying physics of realistic cases, and that
roughly half of the phenomenologically allowed
parameter space can be thought of as
having an induced quartic, and half an induced tadpole.
Note that in the decoupling limit,
all Higgs couplings approach that of the standard model,
and are therefore consistent with  observations.
However, significant deviations from this limit are allowed
by the present data.
If we get too close to the decoupling limit the model becomes tuned,
but the existence of this limit makes it clear why there is a
region where the Higgs has SM-like couplings.
This is conceptually similar to models of Higgs as a 
Nambu-Goldstone boson, where a similar tuned limit 
determines the phenomenology \cite{ArkaniHamed:2001nc,ArkaniHamed:2002qy,Contino:2003ve,Chang:2003un,Agashe:2004rs}.

The induced EWSB mechanism allows the construction of simple SUSY
models that explain the observed 125~GeV Higgs without fine-tuning.
These models are compatible with gauge coupling
constant unification in the simplest possible way, namely that
all new fields come in complete $SU(5)$ muliplets.
The models are also naturally compatible with precision
electroweak constraints.

The simplest $D$-term model is in fact a minor extension of the
`sister Higgs' model described in \Ref{Alves:2012fx}, where the auxiliary
Higgs are the same as sister Higgs fields.
However, we focus on a very different regime of parameters
where the new $SU(2)_S$ gauge group is broken at a scale $\sim 3\TeV$,
naturally suppressing the corrections to precision electroweak
observables as well as gauge coupling unification.
\Ref{Alves:2012fx} instead considered a regime where this breaking
scale is low and focused on the effects of a large $F$-term
generating the quartic $|\Si H|^2$.
In this regime, there is  tension between naturalness
and precision electroweak constraints,
as we will explain below.

The paper is organized as follows.
In \S\ref{sec:supersimplified}
we study further the simplified model of induced EWSB
introduced above.
In \S\ref{sec:Dterms} we discuss models where the auxiliary Higgs quartic
arises from the $D$-term of a new non-Abelian gauge group.
In \S\ref{sec:Fterms} we discuss models where the auxiliary Higgs quartic
arises from an $F$-term and consider a `hybrid' model involving both $F$ and $D$ terms.  Our conclusions are in \S\ref{sec:Conclusions}.  Certain details of the behavior of these models under the renormalization group are reserved for an appendix.

\section{A Simplified Model
\label{sec:supersimplified}}
We now discuss in more detail the simplified model of EWSB defined in the
introduction above.
We remind the reader that this model
consists of two doublets $H$ and $\Si$, with a potential
given in \Eq{Vsupersimp}.
The fully realistic models that we discuss later will have additional
Higgs fields at low energies, but we will see that many aspects of the 
models are similar to the simplified model considered here.

This model has 4 parameters.
Fixing the scale of electroweak symmetry breaking at
$v = 246\GeV$ and the mass
of the lightest CP-even Higgs boson $m_h = 125\GeV$
we have 2 remaining parameters, which we take to be
the quartic coupling $\la_\Si$ and the VEV $f$ of the auxiliary Higgs
field
\begin{align}
\avg{\Si} = \frac{1}{\sqrt{2}}
\begin{pmatrix}
0 \\ f
\end{pmatrix}.
\end{align}
The qualitative features of this parameter space can be understood from
the two limits of this model discussed in the introduction,
and this is what we turn to next.

\subsection{Decoupling Limit}
\begin{figure}[tbp]
\begin{center}
\includegraphics[height=6.cm]{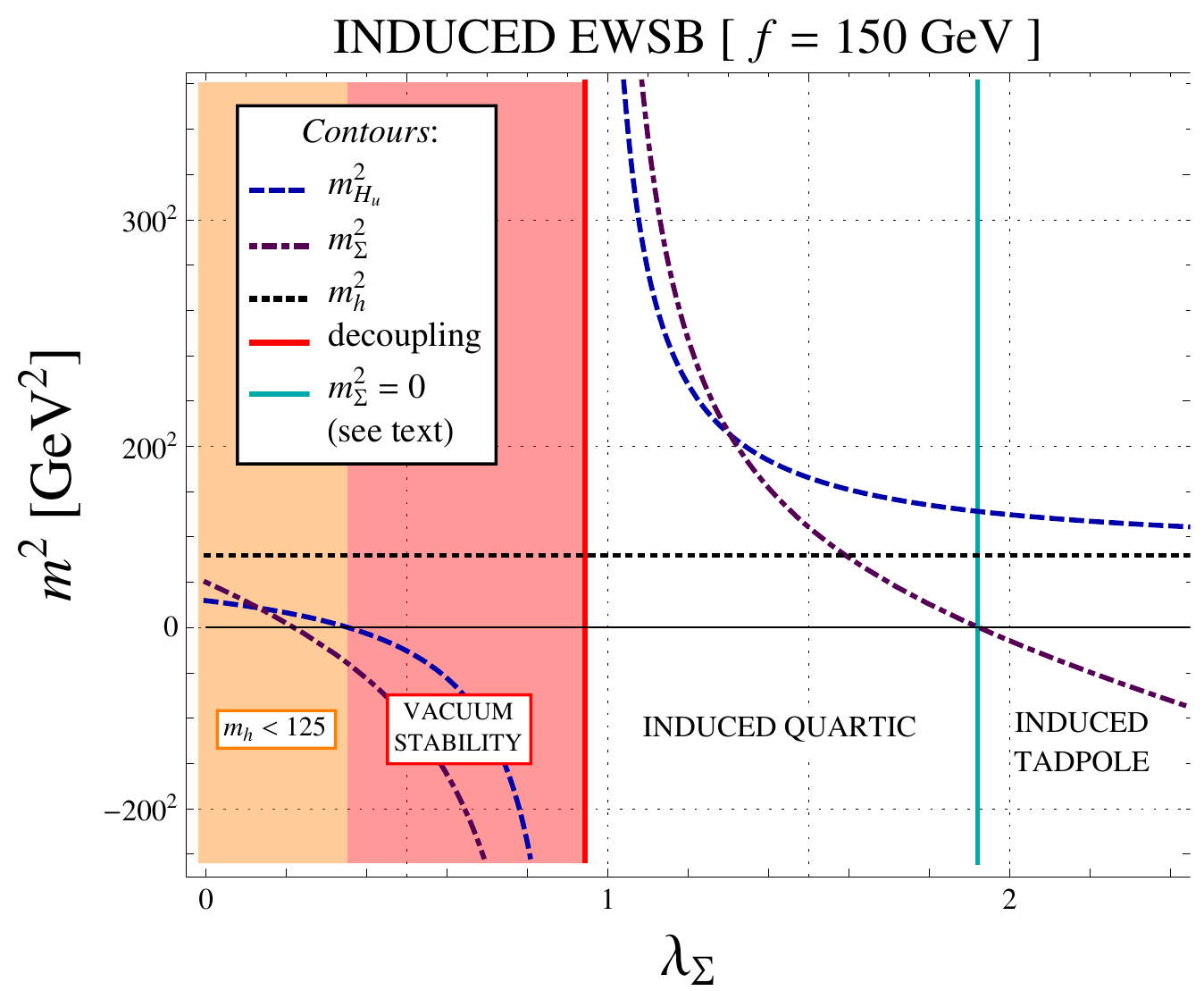}
\caption[]{\small
Higgs mass parameters as a function of the auxiliary Higgs quartic
$\la_\Si$ for $f = 150\GeV$.
The decoupling limit occurs near $\la_\Si = 1$.
$m_\Si^2$ crosses $0$ near $\la_\Si = 2$, indicating a clear transition between
induced quartic and induced tadpole regimes.
\label{fig:lam_soft}
}
\end{center}
\end{figure}

In the decoupling limit, one linear combination of Higgs fields
has a large positive mass-squared term and therefore no VEV.
We can analyze this limit by diagonalizing the quadratic terms in the
potential \Eq{Vsupersimp} by writing
\begin{align}
\begin{pmatrix}
H \\ \Si
\end{pmatrix}
= \begin{pmatrix} s_\ga & c_\ga \\
c_\ga & -s_\ga \\
\end{pmatrix}
\begin{pmatrix}
H_1 \\ H_2 
\end{pmatrix},
\end{align}
where $s_\ga = \sin\ga$, $c_\ga = \cos\ga$.
The potential in terms of these fields is
\begin{align}\eql{eq:effpotential1}
V = m_1^2 |H_1|^2 + m_2^2 |H_2|^2
+ \la_\Si c_\ga^4 |H_1|^4 + \cdots,
\end{align}
where we assume $m_2^2 \gg |m_1^2|$ is the large positive mass-squared
eigenvalue.
Integrating out $H_2$, we then obtain the effective potential for the light
Higgs doublet $H_1$.
The light CP-even Higgs mass is therefore
\begin{align}
m_h^2 = 2 \la_\Si c_\ga^4 v^2.
\end{align}
The mixing angle $\ga$ is determined by the VEVs from the requirement
that $H_2$ have vanishing VEV:
\begin{align}
\tan\ga = \frac{v_h}{f}
\end{align}
so we have
\begin{align}\eql{lambdamin}
\la_\Si = \frac{m_h^2 v^2}{2 f^4}.
\end{align}
This explains the minimum value of $\la_\Si$ in Fig.~\ref{fig:flam_simp}. 
The decoupling limit is approached as $\la_\Si$ approaches the value
\Eq{lambdamin} from above.  
In this limit the soft masses $m_H^2$ and $m_\Si^2$ both grow arbitrarily 
large as is shown in Fig.~\ref{fig:lam_soft}, though with a ratio that 
allows one eigenstate to remain light.  
\Eq{lambdamin} in fact represents a phenomenological lower bound on the quartic: 
for values just {\it below} the decoupling value the Higgs soft mass  \Eq{HiggsSoft} changes sign and the vacuum is consequently destabilized, 
while for very small values of the quartic the Higgs becomes the heavier eigenstate 
of the mass matrix, \Eq{masssupersimp}, indicating the presence of 
scalars lighter than the $125\GeV$ Higgs.

%
%%%%%%%%%%%%%%%%%%%
%%%%%%%%%%%%%%%%%%%
%

In the decoupling limit, the effective theory of electroweak symmetry
breaking is that of a single Higgs doublet, and it is clear that all
Higgs couplings approach those of the standard model.
Because the current data favors a SM-like Higgs, this limit
is compatible with the Higgs data.
Approaching this limit asymptotically, the model becomes fine-tuned because
the heavy mass eigenstate
contributes to the mass of the light mass eigenstate:
\begin{align}
\De m_1^2 \sim \frac{\la_\Si s_\ga^2 c_\ga^2}{16\pi^2} m_2^2.
\end{align}
However, the fine-tuning decreases rapidly away from this limit
and the experimental constraints on the Higgs couplings do not
require the model to be tuned.
This will be made quantitative below.
However, this limit makes it clear that the
model can approximate the standard model, and therefore account for
the current data which are consistent with a SM Higgs.

\subsection{Higgs Phenomenology}
\begin{figure}[tbp]
\begin{center}
\includegraphics[height=6cm]{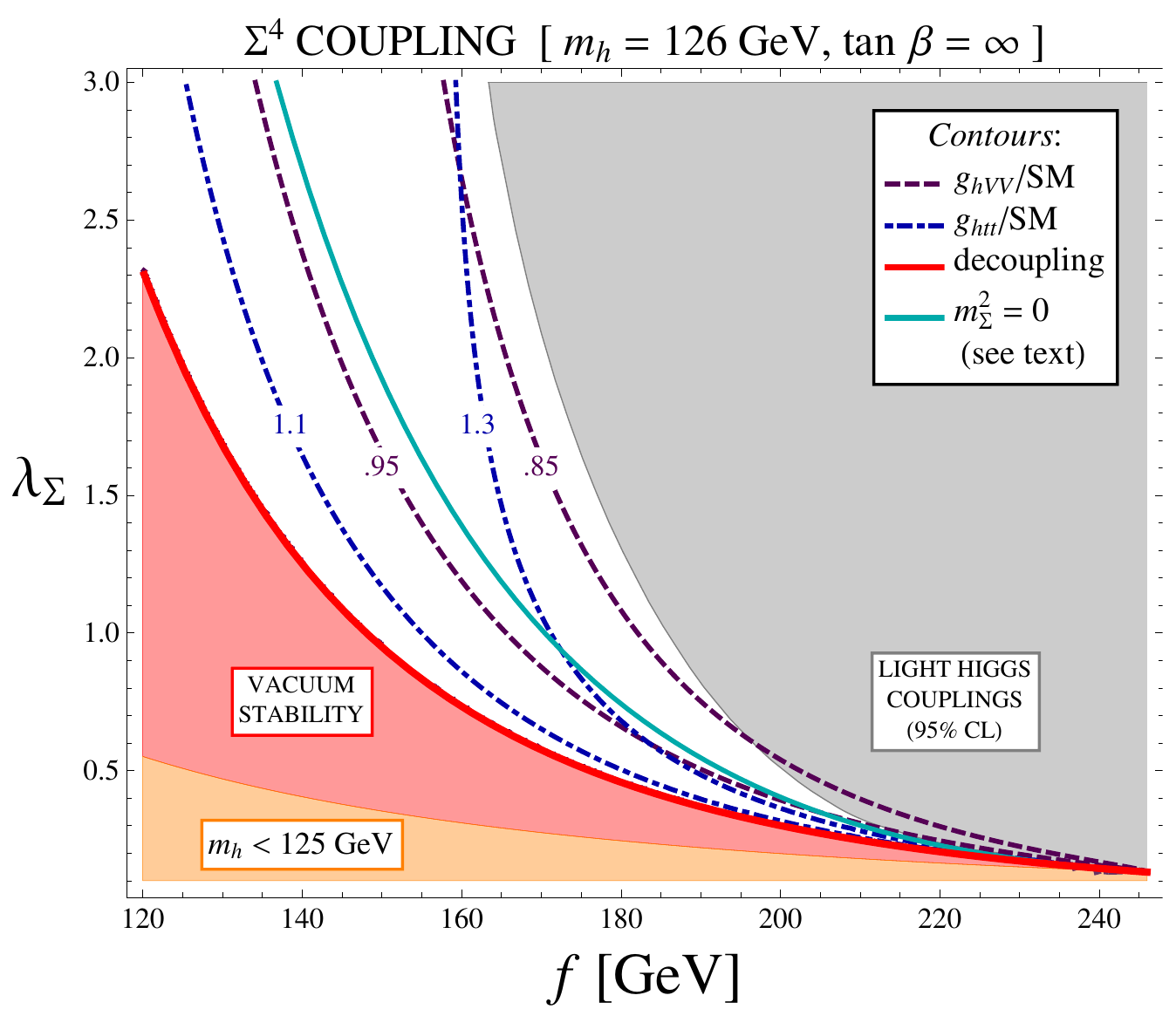} \quad
\includegraphics[height=6cm]{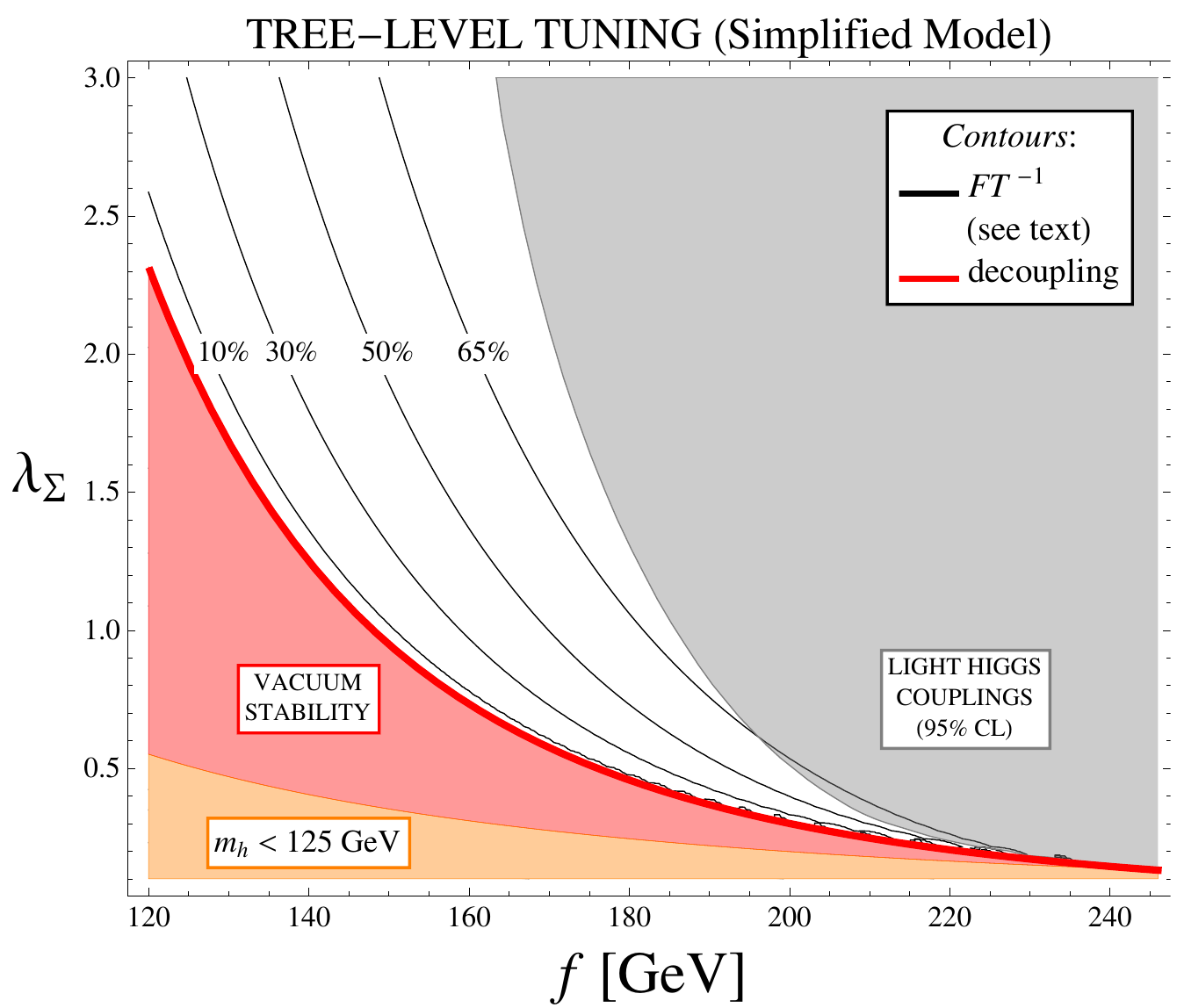}
\caption[]{\small
Parameter space of the simplified two Higgs doublet model describing the
mechanism of induced EWSB.  
{\bf Left:} Couplings of light Higgs couplings to weak gauge bosons %($g_{hVV}$) 
and top quarks. % ($g_{htt}$).  
{\bf Right:} Tree-level tuning of the simplified model, as described in text.
\label{fig:flam_simp}
}
\end{center}
\end{figure}

The results above allow us to easily understand the 
phenomenologically 
allowed parameter range of this model shown in
Fig.~\ref{fig:flam_simp}.  
%
%%%%%%%%%%%%%%%%%%%
%%%%%%%%%%%%%%%%%%%
%
The red and gold regions indicate regions of vacuum instability and
light scalars, respectively.
The grey region is excluded by the measured Higgs couplings at $95\%$
confidence level.
For large values of $f$, significant Higgs mixing is unavoidable
and the model becomes incompatible with measured Higgs couplings.

The red boundary represents the decoupling limit $m_\Si^2 \to \infty$ where all couplings are SM-like.
We emphasize that the light Higgs can still have a finite admixture 
of the $\Si$ fields in this limit.
This should be contrasted to models like the NMSSM where the Higgs gets
a quartic by mixing with a singlet;
in these models, large mixing necessarily implies suppression of the Higgs
couplings to gauge bosons and fermions which is disfavored by LHC and LEP data.

The model becomes fine-tuned as we approach the decoupling limit,
but we do not need to be very close to this limit to satisfy the
phenomenological constraints.
We illustrate this in right panel of Fig.~\ref{fig:flam_simp} by
plotting a standard measure of tuning \cite{Barbieri:1987fn},
namely the inverse of the function
\begin{align}
{\rm FT} = \max \left( \frac{ \partial \log \{v,f,m_h\}}{\partial \log m_{\rm input}^2} \right),
\end{align}
where $m_{\rm input}$ stands for any of the dimensionful parameters defining the potential of \Eq{Vsupersimp}. 
Explicitly the relevant relations are
\begin{align}
\eql{HiggsSoft}
m_H^2 &= m_h^2
\left( 1+ \frac{m_h^2 v_h^2}{2\la_\Sigma f^4 - m_h^2 v^2} \right) \\
\eql{SigmaSoft}
m_\Si^2 &=  m_h^2 v_h^2 \,  
\frac{2 \la_\Sigma f^2- m_h^2 }{2\la_\Sigma f^4 - m_h^2 v^2}  -\la_\Sigma f^2  \\
\eql{KappaSoft}
\kappa^2 &= m_h^2 v_h f \, 
\frac{2 \la_\Sigma f^2- m_h^2 }{2\la_\Sigma f^4 - m_h^2 v^2} .
\end{align}
These become singular in the decoupling limit
$2 \la_\Si f^4 \to m_h^2 v^2$,
accounting for the large tuning there.
When we construct explicit models, there will be other potential sources
of tuning arising from loops involving heavier superpartners,
and we will include these as well.

As we move away from the minimum value of $\la_\Si$, the mass of the heavy
CP-even mass eigenstate first decreases, but then begins to
increase as we enter the regime where
the VEV of the light Higgs field is induced by an effective tadpole.
Because we do not want to tune near the decoupling limit, and we 
do not want to take $\la_\Si$ non-perturbatively large,
we are not really close to either limit.
It is nonetheless interesting to ask whether we are `closer' to
one limit or the other for phenomenologically allowed parameters.

There is no sharp boundary between the two regimes. 
The decoupling limit requires $m_\Si^2 \gg 0$, while the induced
tadpole limit requires $m_\Si^2 < 0$.  A useful definition of the
boundary is thus where $m_\Si^2 = 0$: as can be seen in Fig.~\ref{fig:lam_soft}, the Higgs soft mass beyond this point nearly coincides with its asymptotic value $m_H^2 = m_h^2$ indicating a strongly induced tadpole.  
We illustrate this also in  Fig.~\ref{fig:flam_simp}, and there we see that in
roughly half of the allowed parameter space an induced tadpole
is playing an important role.
In these regions, a physical characteristic of interest is that the Higgs quartic coupling is drastically suppressed relative to its SM value by powers of the small parameter $\ep$
(see \Eqs{Veffsupersimptadpole} and \eq{epsuilondefn}).
The simplified model omits the tree-level contribution to
the quartic from the $SU(2)_W \times U(1)_Y$ $D$-terms,
so the quartic in the simplified model can be viewed as a \emph{correction} to the tree-level mass.  The quartic coupling, as a `smoking gun' of the induced tadpole, will be subject to experimental scrutiny only with significantly more Higgs data than is presently available, but we find it  illustrative to emphasize this feature in the models we describe below.  Thus in realistic cases, we will find it convenient to indicate the importance of the induced tadpole by focussing on regions of the parameter space where the Higgs quartic coupling is smaller than a benchmark value which we will take to be its tree-level value in the MSSM (corresponding to approximately half of the SM value).

We now proceed to a discussion of two separate complete models that can realize the mechanism of induced EWSB with perturbative dynamics.

% =============================================================================
% =============================================================================
\section{A $D$-term Model}
\label{sec:Dterms}
% =============================================================================
% =============================================================================
In this section we consider the case where the large quartic for the
auxiliary Higgs fields arises from the $D$-terms of a new gauge
interaction.
We will see that the simplest versions of this model have a phenomenology
very similar to the simplified model discussed in \S\ref{sec:supersimplified}.
New $D$ terms have been previously considered as a way to generate
large quartics for the Higgs by embedding the Higgs into a multiplet
charged under the new gauge group \cite{Batra:2003nj,Maloney:2004rc}.
However, these models do not preserve gauge coupling unification without
significant additional structure because the Higgs fields are not part
of a complete $SU(5)$ multiplet.
On the other hand, we preserve gauge coupling unification
in our approach simply by taking the auxiliary Higgs fields to be part of a 
complete $SU(5)$ multiplet.

%===========================================================================
\subsection{The Model}
%===========================================================================
We assume the existence of a new gauge interaction with gauge group $SU(2)_S$,
with additional matter fields given in Table~\ref{tab:model}.
\begin{table}[htb]
\begin{center}
\begin{tabular}{c | c | c c c }
 & $SU(2)_S$ & $SU(3)_C$ & $SU(2)_W$ & $U(1)_Y$ \\
 \hline
$\Phi$ & $\square$ & 1 & 1 & 0 \\
$\bar{\Phi} $ &  $\square$ & 1 & 1 & 0  \\ 
$\Si_u$ & $\square$ & 1 & $\square$ & $\frac 12$ \\
$\Si_d$ & $\square$ & 1 & $\square$ & $-\frac 12$ \\
$T$ & $\square$ & $\square$ & 1 & $-\frac 23$ \\
$\bar T$ & $\square$ & $\bar\square$ & 1 & $\frac 23$ \\
\hline
\end{tabular}\caption{\small Field content of the $D$-term model.}\label{tab:model}
\end{center}
\end{table}
Color triplets $T$ and $\bar{T}$ are included so that the model
consists of complete $SU(5)$ multiplets, and therefore preserves gauge
coupling unification.
There are 6 `flavors' of $SU(2)_S$, and therefore it has vanishing
1-loop beta function.
This naturally allows a large range of $SU(2)_S$ gauge coupling
constants at the weak scale.
We will discuss the RG behavior in more detail below.

The $SU(2)_S$ gauge symmetry will be broken at the weak scale by
VEVs of the fields $\Phi$ and $\bar{\Phi}$, as well as the auxiliary
Higgs fields $\Si_{u,d}$.
The $\Si_{u,d}$ VEVs break custodial  symmetry, and therefore 
contribute to the electroweak $T$ parameter.
We will see below that in
order for this to be sufficiently small, we will need 
$u \sim \avg{\Phi},\avg{\bar\Phi} \gg f \sim \avg{\Si_{u,d}}$.
Because we require $f \sim 100\GeV$, we must have 
$u \gsim \mbox{TeV}$ to satisfy precision electroweak constraints.
(This also suppresses the mixing between $SU(2)_W$ and $SU(2)_S$ that
would otherwise ruin the unification prediction of the gauge couplings.)
This creates a potential tuning problem because
the $SU(2)_S$ $D$-term quartic generally gives a tree-level
contribution to the $\Si$ quadratic of order $g_S^2 u^2 \sim \text{TeV}^2$.
We can avoid this problem by assuming that $SU(2)_S$ is broken
along an approximately $D$-flat direction.%
\footnote{This mechanism is also used in models where the Higgs gets a 
contribution to its quartic from new $D$-terms \cite{Batra:2003nj,Maloney:2004rc}.}
We are thus led to introduce the superpotential
\begin{align}
\Delta W =  \la_\Phi S ( \Phi \bar \Phi - w^2 ),
\end{align}
and to impose a $Z_2$ symmetry ensuring approximately
equal soft masses for  $\Phi$ and $\bar \Phi$.   
The potential for these fields then has the form
\begin{align}
V_\Phi= m_\Phi^2 ( \Phi^\dagger \Phi + \bar \Phi^\dagger \bar \Phi ) 
+ \la_\Phi^2 ( \Phi \bar \Phi + w^2 )^2 
+ B_\Phi ( \Phi \bar \Phi + {\rm h.c.})
+V_D,
\end{align}
with $V_D$ the potential arising from $D$ terms of $SU(2)_S$.  
A nonzero VEV is then established provided $B_\Phi > m_\Phi^2 + \la^2 w^2$:
\begin{align}
\langle \Phi \rangle = \frac{1}{\sqrt 2} \begin{pmatrix} 0 \cr u \end{pmatrix} , \quad \langle \bar \Phi \rangle = \frac{1}{\sqrt 2} \begin{pmatrix}  \tilde{u}\cr 0\end{pmatrix},
\end{align}
with
\begin{align}\eql{uformula}
u = \tilde{u} = 
\pm \sqrt{\frac{2 B_\Phi -2 (m_\Phi^2 + \la_\Phi^2 w^2)}{\la_\Phi^2}}.
\end{align}

The quartic for $\Si$ induced from $SU(2)_S$ $D$-terms vanishes in the
SUSY limit, so we must have $m_\Phi^2 \sim g_S^2 u^2$ to avoid a cancellation.
Integrating out the scalar fields in $\Phi$ and $\bar\Phi$ gives
the quartic
\begin{align}
\eql{Dquartic}
\la_\Si \simeq \frac{g_S^2}{8} \left(1 + \frac{ g_S^2(u^2+ \tilde u^2)}{8 m_\Phi^2} \right)^{-1}.
\end{align}
This requires $\la_\Phi \sim g_S$, as can be seen from \Eq{uformula}.
This mechanism therefore requires the existence of two new 
order-1 dimensionless couplings at the weak scale ($g_S$ and $\la_\Phi$).
This raises the issue of Landau poles and perturbative unification,
which will be discussed below.

%===========================================================================
\subsection{A Simple Low-energy Limit}
%===========================================================================
This model can have up to 6 Higgs doublets, $2$ from the MSSM fields $H_{u,d}$ and
$4$ more from $\Si_{u,d}$.
To illustrate the phenomenology, we will focus on a simple limit where
only one of the new Higgs doublets gets a VEV.
The effective theory below the $SU(2)_S$ breaking scale is then a
3-Higgs doublet model.
Specifically, we assume that 
\beq\eql{bidoubletvev}
\avg{\Si_d} = \frac{1}{\sqrt{2}} 
\begin{pmatrix}
0 & 0 \\ 0 & f
\end{pmatrix},
\qquad 
\avg{\Si_u} = 0.
\eeq
For this limit to be exact requires that `$B\mu$ terms' of the form 
$\Si_u \Si_d$ and $\Si_u H_d$ vanish, which is not natural.
However, a mild hierarchy among the SUSY breaking terms can easily
make the simplified limit we study a good first approximation.
In any case, the parameter space of the full model is too large
to study, and we must make some simplifying assumptions to proceed.

With these simplifying assumptions, 
the only terms that are relevant are those that mix $\Si_d$ and
the MSSM Higgs fields.
These are the superpotential terms
\beq\eql{thesuperpotential}
\De W = \mu H_u H_d + \la_u H_u \Si_d \Phi 
+ \bar{\la}_u H_u \Si_d \bar\Phi
\eeq
and the soft SUSY breaking terms in the Higgs potential
\beq
V = m_{H_u}^2 |H_u|^2 + m_{H_d}^2 |H_d|^2 + m_{\Si_d}^2 |\Si_d|^2 
+ \, B\mu H_u H_d + B_u H_u \Si_d  
\eeq
In addition, the theory has the $D$-term quartic from the 
$SU(2)_S \times SU(2)_W \times U(1)_Y$ gauge interactions.

The couplings $\la_u$ and $ \bar \la_u$ contribute the mass terms
\beq\eql{mixingmass}
\De V =  \left( |\la_u|^2 + | \bar{\la}_u |^2 \right)
u^2 
\left( |H_u|^2 + |\Si_d|^2 \right) .
\eeq
Precision electroweak tests and unification require $u \gsim \text{TeV}$,
so in order for these mass terms not to be unnaturally large,
we must take $\la_u, \bar{\la}_u \lsim v/u \sim 0.1$.
This is sufficiently small that $\la_u$, $\bar{\la}_u$ do not
contribute significantly to the quartic.
\Ref{Alves:2012fx} analyzed exactly the same model,
focussing on the Higgs quartic generated by $\la_u$.
However, as discussed here this is fine-tuned at the percent level
if $u \sim \mbox{TeV}$.
\Ref{Alves:2012fx} uses a less stringent bound on the $T$ parameter than we do,
allowing smaller values of $u$.

The couplings $\la_{u}, \bar{\la}_u$
are however relevant because we require $B_u \sim (100\GeV)^2$,
and we expect $B_u$ to be $\la_u u$ times a SUSY breaking mass.
This is consistent with taking $\la_u, \bar{\la}_u \sim 0.1$, which we will assume 
in the following.
We now define
\begin{align}
\tan \be = \frac{v_u}{v_d}, \quad \tan \ga = \frac{v_h}{f}, \quad  v_h = \sqrt{v_u^2+v_d^2}.
\end{align}
The minimization conditions are as follows, 
dropping the now unnecessary subscript from $\Si_d$:
\begin{align}
\eql{eq:mdmin}
m_{H_d}^2 &= B\mu \tan \be  - \frac{1}{2} m_Z^2 \left(s_\ga^2 \cos 2 \be +c_\ga^2 \right) \\
\eql{eq:mumin}
m_{H_u}^2 &= B\mu \cot \be + B_u \frac{\cot \ga}{s_\be} + \frac{1}{2} m_Z^2 \left(s_\ga^2 \cos 2 \be + c_\ga^2\right) -\frac{1}{2}\la_u^2 v^2 c_\ga^2 \\
\eql{eq:msmin}
m_\Si^2  &= B_u s_\be  \tan \ga 
- \frac{1}{2} m_Z^2 \left(s_\ga^2 \cos 2 \be + c_\ga^2 \right)  - \frac{1}{2} g_S^2 v^2 c_\ga^2 - \frac{1}{2}\la_u^2 v^2 s_\be^2 s_\ga^2.
\end{align}
The MSSM formulas are recovered in the limit $s_\ga \to 1$, $B_u \to 0$.
In practice, we solve these for $(B\mu, \, B_u, \, m_{\Si}^2)$ respectively.

The $SU(2)_W \times U(1)_Y$ $D$-term quartics necessarily give a small
contribution to a $125\GeV$ Higgs mass, so it is a good approximation
to neglect them.
We also assume that stop loops do not contribute a large quartic.
In our final numerical results we include these effects, but we can obtain
a good analytic approximation by dropping them.
In this approximation, the mass matrix for the $CP$-even neutral Higgs
bosons is
\beq
\!\!
\mathcal M^2 =
\begin{pmatrix}
m_{H_u}^2 & -\frac{v_d}{v_u}  m_{H_d}^2 & -\frac{1}{v_u f} (m_{H_u}^2 v_u^2- m_{H_d}^2 v_d^2) \cr
-\frac{v_d}{v_u}  m_{H_d}^2 & m_{H_d}^2 & 0 \cr
-\frac{1}{v_u f}(m_{H_u}^2 v_u^2- m_{H_d}^2 v_d^2) & 0 & \frac{1}{f^2} (m_{H_u}^2 v_u^2- m_{H_d}^2 v_d^2)+ \frac 14 g_S^2 f^2
\end{pmatrix}.
\eeq
We have set $\la_u, \bar{\la}_u \simeq 0$ as explained above.
Assuming large $\tan\be$ ($v_d \ll v_u, f$)
we obtain an illustrative upper bound on the lowest eigenvalue:
\beq
m_h^2 < m_Z^2 \cos^2 2 \ga 
+ \sfrac{1}{4} g_S^2 v^2 \cos^4 \ga.
\eeq
This result is similar to what is obtained in models where new $D$-terms
modify the Higgs quartic, but in this case the effect comes from
contributions to the auxiliary Higgs quartic while allowing the physical Higgs quartic to remain suppressed.

Below the $SU(2)_S$ breaking scale, this simplified limit has 6 parameters:
\beq
m_{H_u}^2,\ 
m_{H_d}^2,\ 
B\mu,\ 
m_{\Si_d}^2,\
B_u,\ 
\la_\Si.
\eeq
We write 
\beq
\la_\Si = \frac{g_{S\gap{\rm eff}}^2}{8},
\eeq
where $g_{S\gap{\rm eff}}$ and the $SU(2)_S$ gauge coupling
coincide  in the limit $m_\Phi^2 \to \infty$ (see \Eq{Dquartic}).
After fixing $v$ and the mass of the lightest $CP$ even mass eigenstate, 
there are 4 free parameters.
We take two of these to be $f$ and $g_{S\gap{\rm eff}}$, and scan over the rest.
We will see that the parameter space in the plane of $f$ and 
$g_{S\gap{\rm eff}}$ reproduces the main features of the simplified
model discussed in \S\ref{sec:supersimplified}.
In particular, the lower bound on $g_{S\gap{\rm eff}}$ corresponds to the
limit where one linear combination of Higgs doublets decouples,
and this line is determined to a good approximation by \Eq{lambdamin}
for the simplified model.

\begin{figure}[hbt]
\begin{center}
\includegraphics[height=6cm]{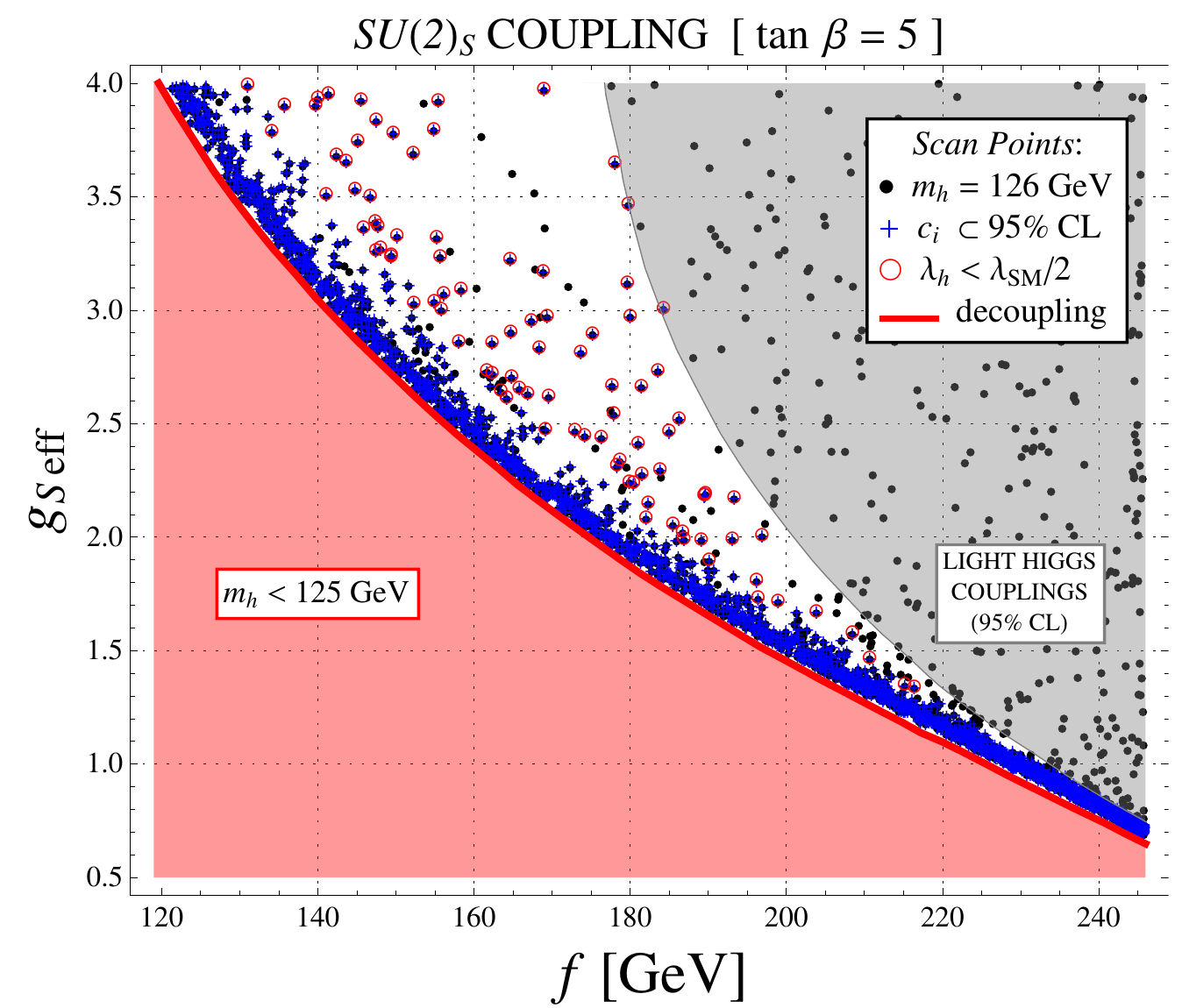} 
\caption{\small Scan of the parameter space for a model where auxiliary Higgs 
quartics arise from non-decoupling $D$ terms.
All points have $m_h \simeq 125\GeV$.
A blue cross is added if the Higgs couplings are compatible with experimental values, and a red circle is added to surviving points if the physical Higgs has a quartic of order half its SM value, indicating regions where the induced tadpole 
is important.   
}
\label{fig:Dterms_tb5}
\end{center}
\end{figure}

The couplings of the light Higgs to the $W/Z$, the top quark, and the bottom
quark relative to their SM values are given by 
\begin{align}
\eql{cV}
c_V &=
\braket{\Si_d}{h} \cos\ga
+\left( \langle H_u^0 | h\rangle \sin\be + \langle H_d |h \rangle \cos\be \right)
 \sin \ga   
\\
\eql{ct}
c_t  &= \frac{ \langle H_u^0 | h\rangle}{\sin \ga \sin \be} \\
\eql{cb}
c_b  &= \frac{ \langle H_d^0 | h\rangle}{\sin \ga \cos \be}.
\end{align}
We show the allowed region for a benchmark case $\tan\be = 5$ in Fig.~\ref{fig:Dterms_tb5}, now including the subleading SM quartic contributions.
As in the simplified model, the lower boundary corresponds to the decoupling
limit for the auxiliary Higgs bosons, and near this limit the light Higgs
is SM-like.
The grey region is excluded by Higgs coupling constraints, determined as in \cite{AGrev}.
This is the region of small $\sin\ga$, which suppresses $c_V$
and enhances $c_t$.
Fig.~\ref{fig:Dterms_tb5} also shows the allowed points where $\la_h$ is smaller
than half of its SM value.
In such theories, the Higgs cubic and quartic self-couplings are highly
suppressed compared the  SM, which can be measurable in
future experiments or with large integrated luminosity via enhanced di-Higgs production at the high-energy LHC \cite{Klute:2012pu, Dolan:2012rv, Goertz:2013kp}.
This also has conceptual importance, since it corresponds to the regime
where an induced tadpole is important, as discussed in \S\ref{sec:supersimplified}.
The tree-level MSSM gives a quartic that is about half of the SM
value (for large $\tan\be$), so these points are also very far from the
usual SUSY solutions, all of which strive to obtain a Higgs quartic close to 
SM value.

%%%%%%%%%%%%%%%%%%%%%%%%%%%%%
%%%%%%%%%%%%%%%%%%%%%%%%%%%%%
%%%%%%%%%%%%%%%%%%%%%%%%%%%%%
%%%%%%%%%%%%%%%%%%%%%%%%%%%%%

%===========================================================================
\subsection{Electroweak Precision Tests}
%===========================================================================
We turn now to the issue of corrections to electroweak precision tests.
The strongest constraints are oblique radiative corrections,
which we treat using 
the formalism of \cite{Peskin:1990zt,Peskin:1991sw}. 

There are two main effects. 
The first comes from the $SU(2)_S$ breaking threshold at the scale $g_S u$.
The VEVs of the auxiliary Higgs
fields break both $SU(2)_S$ and the electroweak group.
This breaks custodial symmetry, and therefore gives a positive tree-level
contribution to the $T$-parameter.
The other main effect comes from loops of the auxiliary Higgs
fields.
As we explain below, 
for large $g_S$ the auxiliary Higgs fields contribute positively to the
$S$ parameter like a heavy Higgs doublet, while giving a negligible contribution
to the $T$-parameter.

These two effects partially offset each other, in the sense that
a positive $S$ parameter
allows a larger positive $T$ parameter, allowing lower values of the 
$SU(2)_S$ breaking scale than would be allowed if only one of these
effects was present.
This in turn reduces the leading source of fine-tuning in this model.

\paragraph{$SU(2)_S$ breaking:}
The $SU(2)_S$ gauge bosons mix with the electroweak gauge bosons
due to VEVs of the auxiliary Higgs bosons.
A $T$-parameter  arises because the $W_S^{1,2}$ gauge bosons do not mix
with the $W^\pm$, while the $W^3_S$ does mix with the neutral
SM gauge bosons.

The mixing of the neutral gauge bosons arises from the VEVs
of the axiliary Higgs fields via the kinetic term.
For illustration we will focus on the case of a single auxiliary
Higgs, but the generalization to additional Higgs is straightforward.
We have
\begin{align}
\Delta \lag = \left| - i g \frac{\tau_3}{2} W_{\mu}^3  \Si_d
+ i g' B_\mu \frac 12 \Si_d 
- i g_S W^3_{S \gap \mu} \Si_d \tau_3  \right|^2
\end{align}
and a mass matrix for the neutral gauge bosons in the basis
$(W_\mu^3, B_\mu, W_{S\gap\mu}^3)$ given by
\begin{align}
\mathcal M^2 = \frac{1}{4}\begin{pmatrix}
g^2 v^2 & - g g' v^2 & - g g_S f^2 \cr
- g g' v^2 &  g'^2 v^2 &  g' g_S f^2 \cr
 - g g_S f^2 &  g' g_S f^2 &  g_S^2 \left(f^2 + u^2 + \tilde{u}^2\right) 
\end{pmatrix}.
\end{align}
Integrating out the heavy neutral gauge boson, the leading correction 
to the effective theory is given by
\beq
\De\scr{L}_\text{eff} = -\frac{1}{8} \frac{(g^2 + g'^2) f^4}{u^2 + \tilde{u}^2}
Z^\mu Z_\mu
\eeq
where $Z_\mu = (g W^3_\mu - g' B_\mu) / \sqrt{g^2 + g'^2}$.
This is a correction to the $Z$ mass without a corresponding 
correction to the $W$ mass, which gives
\begin{align}
\Delta T = -\frac{1}{\al} \frac{\delta m_{Z}^2}{m_Z^2} 
= \frac{1}{\al}  \frac{f^4}{\left(u^2 + \tilde{u}^2\right) v^2 }.
\end{align}

\begin{figure}[tbt]
\begin{center}
		\begin{tikzpicture}[line width=1.0 pt, scale=1.2] 
			%\draw[color=white] (-1.7,-1.7) rectangle (1.7,1.7);
			\draw[vector] (-2.4,0) -- (-.9,0);
			\draw[scalarnoarrow] (-1,0) -- (-1.75,.6);
			\draw[scalarnoarrow] (-1,0) -- (-0.25,.6);
			\draw[scalarnoarrow] (1,0) -- (1.75,.60);
			\draw[scalarnoarrow] (1,0) -- (0.25,.60);

		        \draw[fermionnoarrow][fill=black] (-.87,0) arc (-360:135:.13);
		         \draw[fermionnoarrow][fill=black] (1.13,0) arc (-360:135:.13);
		%	\draw[fermion] (135:1) arc (135:45:1);
		%	\draw[fermion] (45:1) arc (45:0:1); 
					\draw[vector] (1.1,0) -- (2.4,0);
			\draw[vector] (-.87,0) -- (.99,0);
			%
			%		 	 \node at (0,-1.4) {$H^-$};
           \node at (-2.8,0) {\small{$W^3_{\mu}$}};
           \node at (-1.8,.9) {\small{$f$}};
           \node at (-0.2,.9) {\small{$f$}};
           \node at (1.8,.9) {\small{$f$}};
           \node at (0.3,.9) {\small{$f$}};
              \node at (2.8,0) {\small{$B_{\mu}$}};
                   \node at (0,-.5) {\small{$W^3_{S,\mu}$}};
		\end{tikzpicture}
\caption{\small Tree-level diagram contributing to the $S$ parameter.}
	\label{fig:Sparameter}
	\end{center}
%\end{figure}
\end{figure}
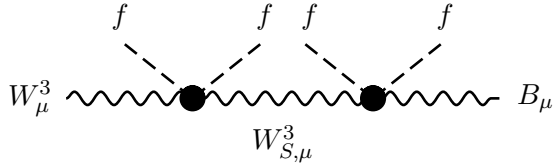

%
%\paragraph{The $S$ Parameter:}
The leading order contribution to the $S$ parameter comes from mixing between 
$W^3_\mu$ and $B_\mu$ generated by mixing with an intermediate $SU(2)_S$ boson 
$W_{S\gap\mu}^3$.  
From the tree-level diagram in Fig.~\ref{fig:Sparameter}
we can integrate out $W^3_S$ with $m^2_{W^3_S}=g_S^2(u^2+\tilde{u}^2+f^2)/4$ to arrive at the form factor 
\begin{align}
\Pi_{3B}(p^2) &= -\frac{(g g_S f^2)(g' g_S f^2)}{4\,(p^2-m^2_{W^3_S})} \nonumber \\
&=\frac{g g' f^4}{u^2+\tilde{u}^2+f^2}\left[1+\frac{4 p^2}{g_S^2(u^2+\tilde{u}^2+f^2)}+\mathcal{O}(p^4)\right].
\end{align}
Thus for the $S$ parameter,
\begin{align}
S=\frac{16 \pi}{g g'} \Pi'_{3B}(0) 
\simeq 64 \pi \frac{ f^4}{g_S^2 (u^2+\tilde{u}^2+f^2)^2}.
%=\frac{\pi\,\alpha\,f^4}{s_W\,c_W\,g_s^2\,(u^2+\tilde{u}^2+f^2)}.
\end{align}
Taking conservative values (see below)
$g_S = 1$, $\tan \ga = 2$, and  $u, \tilde{u}\simeq\mbox{TeV}$, 
we find a completely negligible $S$ parameter,
$S \simeq 5 \times 10^{-3}$.

\paragraph{Auxiliary Higgs bosons:}
In the limit where $g_S$ is large, the auxiliary Higgs fields act as
a heavy electroweak breaking sector, and there is a danger of a
large contribution to the $S$ and $T$ parameters.
In this limit we can integrate out the auxiliary Higgs fields
and write an effective theory where
electroweak symmetry is nonlinearly realized.
This effective theory will contain explicit $W$ and $Z$ mass terms proportional 
to $f$.
This effective Lagrangian will also contain a 4-derivative operator corresponding to the
$S$-parameter.
This is dimensionless and hence depends only logarithmically on the mass
of the heavy states:
\beq\eql{Sloop}
\De S \simeq \frac{1}{12\pi} \ln\frac{m_\si^2}{m_h^2},
\eeq
where $m_\si$ is the mass of the heavy auxiliary Higgs bosons.
This expression is correct only in an approximation where the 
logarithm dominates, but this is the only limit where such a contribution
is large.
The low-energy effective theory also contains vector boson mass terms
that violate custodial symmetry, but these are proportional to $f^2$,
and hence the $T$-parameter is given by
\beq\eql{Tloop}
\De T \simeq -\frac{3}{16\pi c_W^2} \ln \frac{m_\si^2}{m_h^2}
\times \frac{f^2}{v^2}.
\eeq
This gives a small contribution to the $T$-parameter unless $f \simeq v$.

The other limit that is easy to understand is the decoupling limit,
which corresponds to taking $g_S$ as small as possible.
In this limit the auxiliary Higgs fields are again heavy, but
their masses are nearly electroweak-preserving.
Therefore there is a contribution to the $S$ and $T$ parameter
that vanish as we approach this limit.
Explicitly for the $S$ parameter, we have
\beq\eql{Sdecoupling}
\De  S = \frac{g_S^2}{96 m_{\sigma}^2} \frac{f^2 v_h^2}{v^2}
\eeq
The corrections are therefore negligible in the decoupling  limit.

%%%%%%%%%%%%%%%%%%%%%%%%%%%%%
%%%%%%%%%%%%%%%%%%%%%%%%%%%%%
\begin{figure}[tbt]
\begin{center}
\includegraphics[height=6cm]{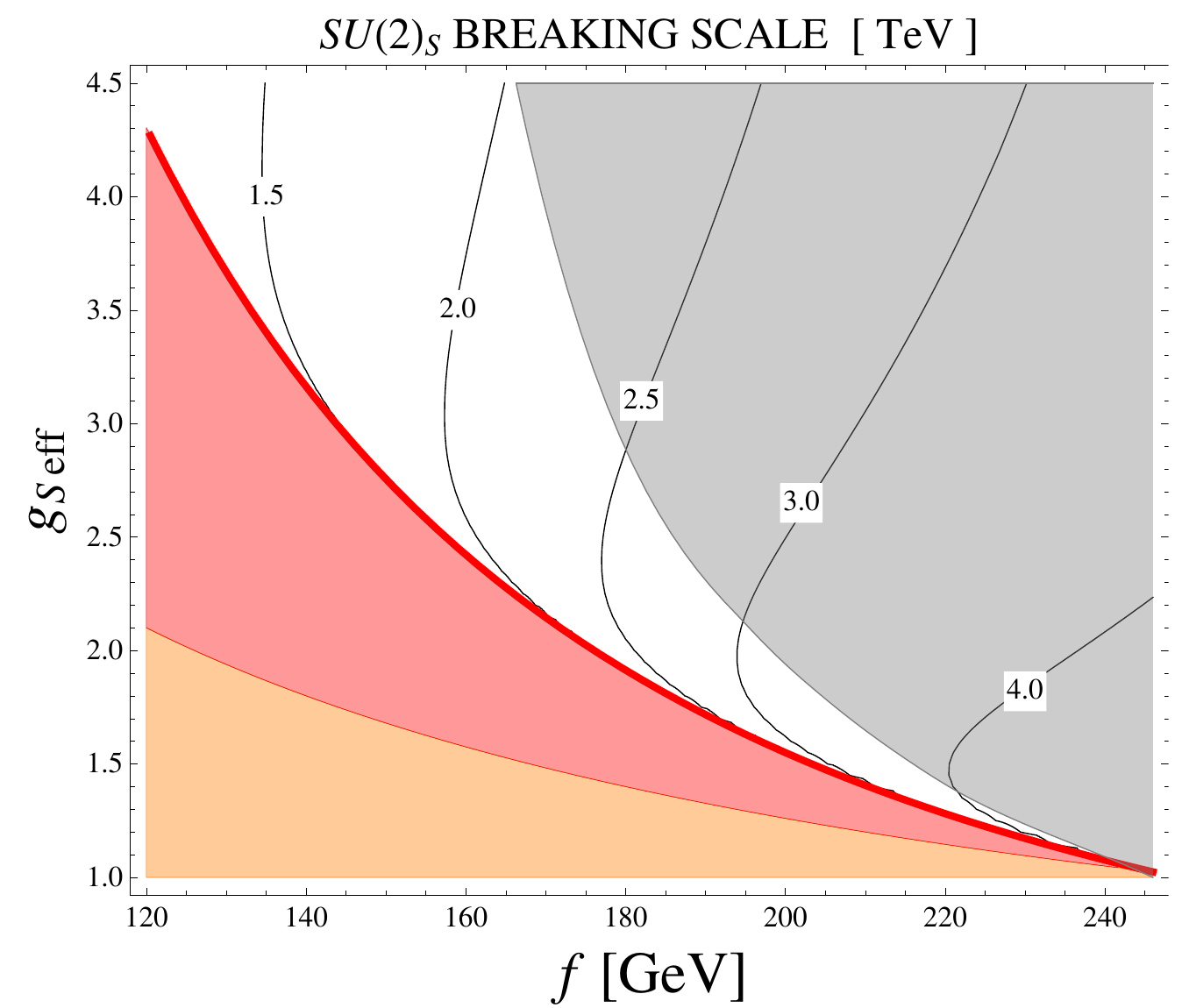} 
\caption{\small 
Bounds on the $SU(2)_S$ breaking scale $u$ from precision electroweak
constraints.}
\label{fig:PEWplot}
\end{center}
\end{figure}
%%%%%%%%%%%%%%%%%%%%%%%%%%%%%
%%%%%%%%%%%%%%%%%%%%%%%%%%%%%

To include these effects, we include the full perturbative
contribution from the Higgs sector using the results of \cite{Haber:2010bw}.
We impose the $95\%$ confidence level $S$-$T$ constraint from \cite{Baak:2012kk}
with $U = 0$.
We find that the $S$ parameter contribution from the auxiliary Higgs
bosons is never so large that it cannot be offset by a positive
$T$ contribution.
The result can therefore
be given as a constraint on the $SU(2)_S$
breaking scale $u$, and is shown in Fig.~\ref{fig:PEWplot}.
We see that this scale must be at least $2$--$3\TeV$ in
most of the parameter space.

%===========================================================================
\subsection{Naturalness}
\begin{figure}[tbt]
\begin{center}
\includegraphics[height=6cm]{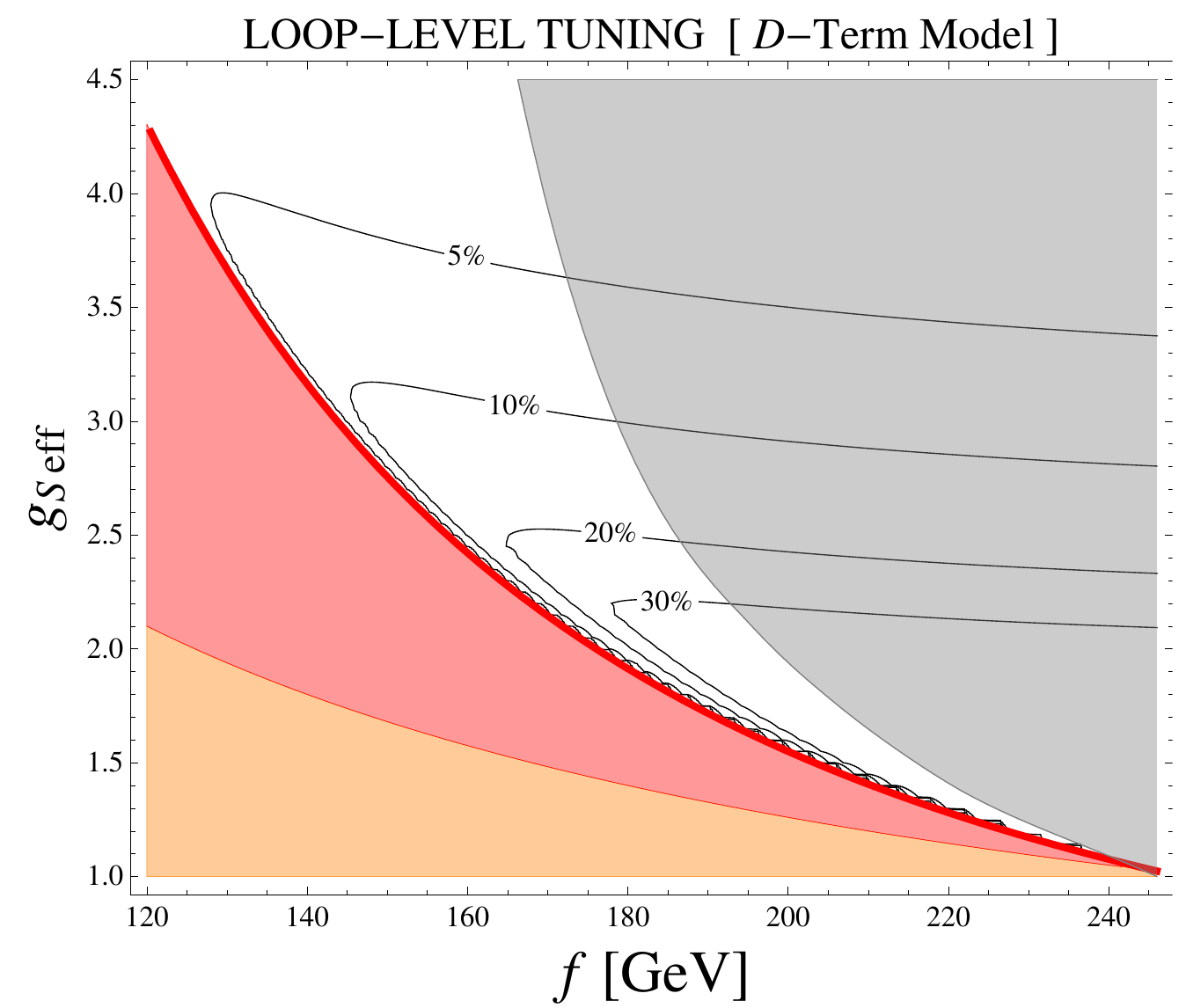} 
\caption{\small Fine tuning in the simplified model with non-decoupling $D$ terms.  Tree-level contributions dominate near the decoupling (bold red) contour; two-loop contributions from the heavy scalar $\Phi$ dominate at increased values of $f, g_S$ where  $u, \tilde u \gg v$ and $m_\Phi^2 \gg m_\Si^2$.  Excluded regions correspond to those described in Fig.~\ref{fig:flam_simp}.}
\label{fig:DtermTuning}
\end{center}
\end{figure}

%===========================================================================
We now discuss the question of naturalness in this model.
We have seen above that the $SU(2)_S$ breaking scale is required
to be in the TeV range, while the auxiliary Higgs fields
(which are also charged under $SU(2)_S$) must have VEVs near the
$100\GeV$ scale.
Loop corrections can potentially destabilize this little hierarchy
and give rise to fine-tuning.
From \Eq{Dquartic} we know that the mass-squared of $\Phi$ must be large, of order $g_S^2 u^2$, in order to have an unsuppressed quartic coupling for the $\Si$ fields.
The only large coupling between the $\Phi$ and $\Si$ fields is the $SU(2)_S$
gauge coupling, and so the leading correction to the $\Si$ mass arises at
2 loops.
We have  \cite{Martin:1993zk}
\begin{align}
\eql{Sigma2loop}
\delta m_\Si^2 = \frac{3g_S^4}{(16\pi^2)^2}\, m_{\Phi}^2 \log \left(\frac{\La^2}{m_\Phi^2}\right),
\end{align}
with $\La$ the mediation scale.
We illustrate the size of the resulting tuning within the context of the simplified model (i.e. treating SM quartics as negligible) in Fig.~\ref{fig:DtermTuning}.
%%%%%%%%%%%%%%%%%%%%%%%%%%%%%
%%%%%%%%%%%%%%%%%%%%%%%%%%%%%
%%%%%%%%%%%%%%%%%%%%%%%%%%%%%
%%%%%%%%%%%%%%%%%%%%%%%%%%%%%
We set $\La = 50\TeV$ and choose $m_\Phi^2 = g_S^2 u^2/4$
so as to consider only parameters where the auxiliary Higgs quartic
is unsuppressed.
We also enforce the precision electroweak constraints as described above.
This puts a lower bound on $u$ and therefore drives the fine-tuning.
We see that the tuning becomes large in the decoupling limit,
as well as the limit of large $g_S$, where the 2-loop effect
\Eq{Sigma2loop} is enhanced.
The tuning is less than $10\%$ over \emph{most} of the allowed parameter
space.
This is a significant difference between this mechanism and other
proposed solutions to the Higgs tuning problem, which have locally
small tuning only in a very specific range of parameters.
In this model, the only requirement is that the $SU(2)_S$ breaking scale
be close to the smallest phenomenologically allowed value.
There are no other large mass hierarchies required in this model,
and we conclude that there is no significant tuning in this model
subject to only very mild restrictions of the parameters.

%===========================================================================
\subsection{Unification and Landau Poles}
%===========================================================================
The fact that the strong and electroweak
guage couplings unify at high scales in the MSSM is
a possible hint for the existence of SUSY in nature, and it is important
to know to what extent unification is  naturally incorporated into 
extensions of the MSSM such as the one we are considering.
The extra fields in our model come in complete $SU(5)$ multiplets,
but this by itself is not sufficient to ensure unification.

%%%%%%%%%%%%%%%%%%%%%%%%%%%%%
%%%%%%%%%%%%%%%%%%%%%%%%%%%%%
\begin{figure}[tbt]
\begin{center}
\includegraphics[height=6cm]{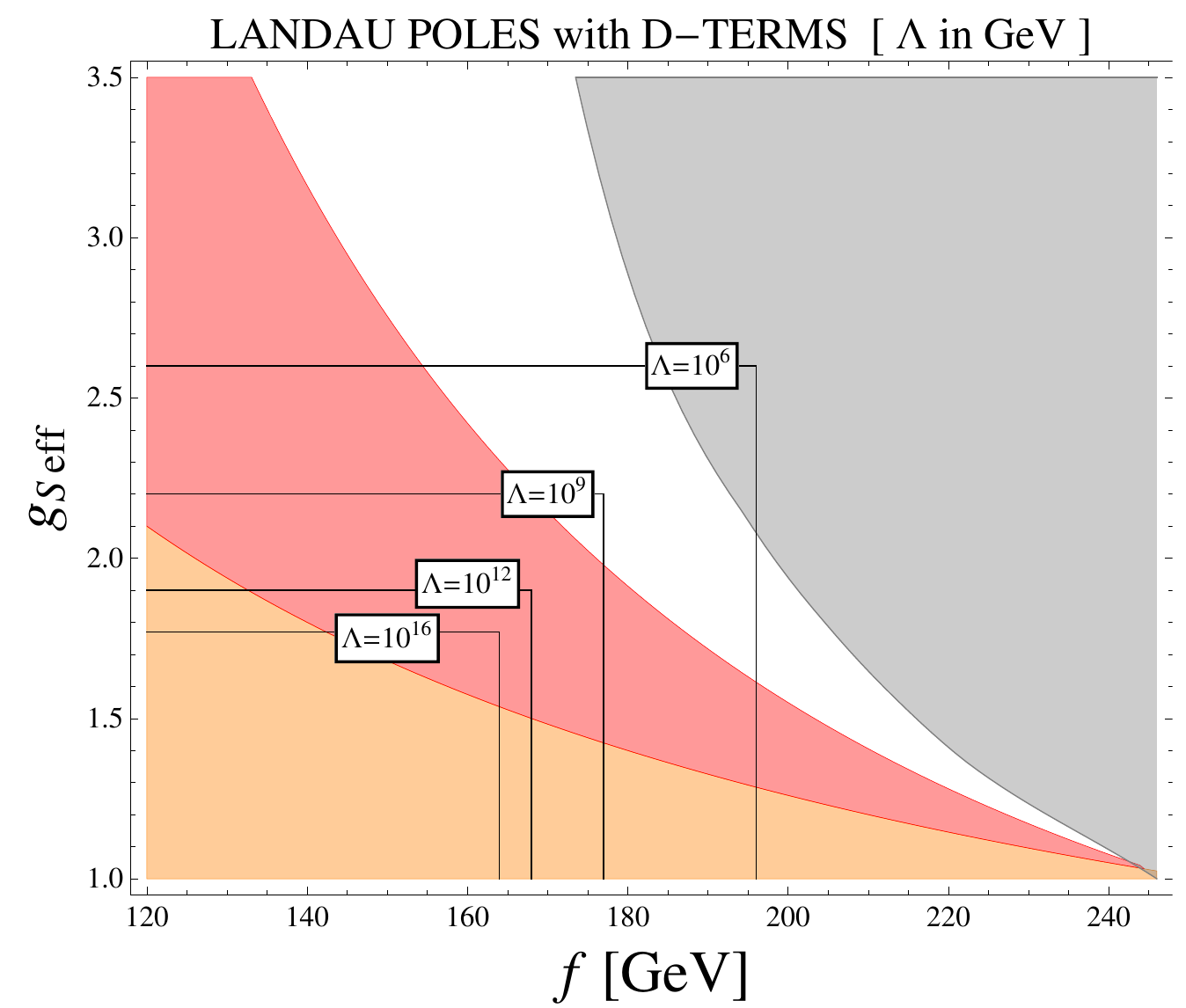} 
\caption{\small Landau poles in the parameter space of the $D$-term model. 
The unshaded area is the allowed region,
as in Fig.~\ref{fig:flam_simp}.}
\label{fig:DDrunning}
\end{center}
\end{figure}
%%%%%%%%%%%%%%%%%%%%%%%%%%%%%
%%%%%%%%%%%%%%%%%%%%%%%%%%%%%

One issue is that the
VEVs of the auxiliary Higgs fields break both $SU(2)_S$ and $SU(2)_W$,
and therefore mix the gauge bosons from these groups and change the
value of the measured weak gauge coupling.
However, the correction is suppressed because of the large $SU(2)_S$
breaking from the VEVs of the fields $\Phi$ and $\bar\Phi$.
The correction to the measured low-energy gauge coupling is of order 
$f^4/u^4$, which is negligibly
small once we impose the $T$ paramter constraint.

Another issue is that 
this model requires that the couplings $g_S$ and $\la_\Phi$
be order unity at the TeV scale.
Neither coupling is asymptotically free, so there is a danger that the
running couplings become large below the GUT scale,
potentially ruining gauge coupling unification.
Also, large values of $f$ require a larger value of $y_t$ at the
weak scale, so there is the danger of a Landau pole for this coupling
as well.
The maximum scale of perturbativity as a function of the parameters
is indicated in Fig.~\ref{fig:DDrunning}.
As we expect, we find that a Landau pole appears at lower scales
when either $g_S$ or $f$ gets large.

There is a Landau pole well below the GUT scale for all allowed
parameters, and therefore this theory requires UV completion below
that  scale.
Such UV completions can naturally preserve unification if the
new physics comes in complete $SU(5)$ multiplets.
An important point in the present model is that
all fields with the large $g_S$
coupling 
satisfy exactly this GUT criterion.
This makes it simple to have new physics that avoids the Landau
pole in $g_S$ while
preserving gauge coupling unification.
As a simple illustrative example of this point, we consider a model where $SU(2)_S$ is embedded
into a larger gauge group $SU(3)_{S'}$ at a higher scale $M$.
The particle content is given in Table~\ref{tab:DXmodel}.
In order to break $SU(3)_{S'} \to SU(2)_S$ we need the additional
superpotential terms
\beq\eql{DXW}
\De W \sim \la' S' (\bar{\Phi}' \Phi' - M^2)
+ F (\bar{\De} \Phi') + \bar{F} (\De \bar{\Phi}'),
\eeq
where $S'$ is a singlet.
We assume that the scale $M$ is above the TeV scale, so the
theory is approximately supersymmetric at the scale $M$.
The $F$- and $D$-flat conditions fix the VEVs
\beq
\avg{S'} = 0,
\qquad
\avg{\Phi'} = \avg{\bar{\Phi}'} = 
\begin{pmatrix} 0 \\ 0 \\ M \end{pmatrix}.
\eeq
The second and third terms in \Eq{DXW}
generate masses of the form $\bar{F} \De_3$ and $F \bar{\De}_3$, with the additional components of the $\Delta$ fields matching onto $\Sigma_{u,d}$.
The theory has an accidental $U(1)$ global symmetry under which $\Phi'$
and $\bar{\Phi}'$ have opposite charges, and there is therefore a massless singlet
Goldstone chiral multiplet.
All other fields get masses of order $M$, and the low energy
theory is the model discussed above.
The additional Goldstone multiplet can get masses from higher
dimension operators, and is harmless.

\begin{table}[ttb]
\begin{center}
\begin{tabular}{c | c  c}
 & $SU(3)_{S'}$ & $SU(5)_{\rm SM}$ \\
 \hline
$\Phi, \Phi'$ & $\square$ & 1 \\
$\bar{\Phi}, \bar{\Phi}'$ &  $\bar\square$ & 1 \\ 
$\De$ & $\square$ & $\square$ \\
$\bar{\De}$ & $\bar\square$ & $\bar\square$ \\
$F$ & 1 & $\square$ \\
$\bar{F}$ & 1 & $\square$ \\
\hline
\end{tabular}\caption{\small Field content of the extended $D$-term model.
The fields are in complete multiplets of 
$SU(5)_{\rm SM} \supset SU(3)_C \times SU(2)_W \times U(1)_Y$.}
\label{tab:DXmodel}
\end{center}
\end{table}

\begin{figure}[tbt]
\begin{center}
\includegraphics[height=6cm]{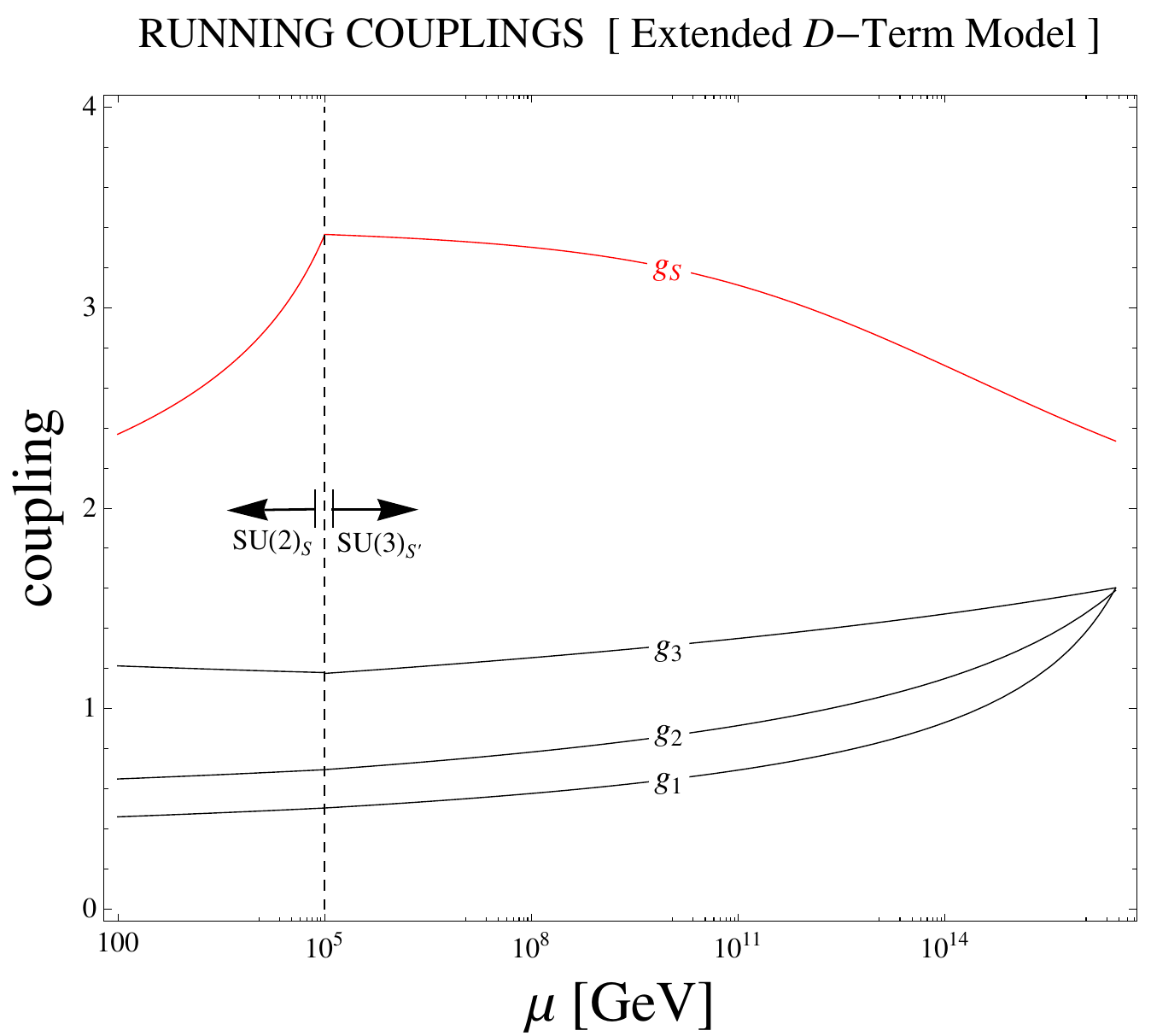} 
\caption{\small Running couplings and their matching at the scale $M \sim 10^5 \, {\rm GeV}$ where $SU(3)_{S'}$ is broken to $SU(2)_S$.  We take $f=160 \, {\rm GeV}$ such that the top coupling remains perturbative and the  tuning required is $\mathcal O( 15 \%)$.}
\label{fig:extended}
\end{center}
\end{figure}

Above the scale $M$, the $SU(3)_{S'}$ gauge group has 7 flavors,
and is therefore asymptotically free.
In fact it is in the conformal window 
\cite{Seiberg:1994pq,Intriligator:1995au,Intriligator:2003jj},
and we expect that there is an IR-stable fixed point for the gauge coupling
and Yukawa coupling $\la_\Phi$.
It is therefore simple to find an RG trajectory where $g_S$
and $\la_\Phi$ are perturbative at the GUT scale, and run toward
the fixed point in the IR.
One possibility is that the scale $M$ is near but above the
scale $\Lambda_{S'}$ where the $SU(3)_{S'}$ coupling becomes strong.
This requires a coincidence of scales.
We can avoid this coincidence if the $SU(3)_{S'}$ coupling
becomes strong and flows to its fixed point above the scale $M$.
The operator $\bar{\Phi}' \Phi'$ has dimension $\frac{12}{7}$, and
the coupling $\la'$ in \Eq{DXW} therefore has a large anomalous
dimension and becomes strong at a scale
\beq
\La_{\la'} \sim \La_{S'} \left( \frac{\la'(\La_{S'})}{4\pi} \right)^{7/2}.
\eeq
For $\la'(\La_{S'}) \sim 0.1$ we obtain
$\La_{\la'} / \La_{S'} \sim 10^{-7}$, so the coupling $\la'$
can naturally remain perturbative at scales far below
the scale where the $SU(3)_{S'}$ fixed point is reached;
the $F(\bar{\De} \Phi')$ and $\bar{F} (\De \bar{\Phi}')$
couplings behave in the same way.
We therefore consider the case where the superpotential
couplings break the $SU(3)_{S'}$ at scales below $\La_{S'}$
but above the scale where $\la'$ gets strong.
In this regime, the $S'$ term in the superpotential does not have
a large anomalous dimension, and the $SU(3)_{S'}$ breaking
scale is given by
\beq
\tilde{M} \sim M^{7/6} \La_{S'}^{-1/6}.
\eeq
This is very insensitive to the scale $\La_{S'}$,
so the breaking scale is still essentially set by $M$.
The bottom line is that there is a large regime of parameters
where the RG trajectory reaches the strong $SU(3)_{S'}$
fixed point, and the theory breaks to the $SU(2)_S$ theory
at a scale set by the parameter $M$.
This theory is strongly coupled, so we can naturally obtain
a large value of the $SU(2)_S$ gauge coupling at the weak scale.  
The salient features of this scenario are sketched in Fig.~\ref{fig:extended}.
%%%%%%%%%%%%%%%%%%%%%%%%%%%%%
%%%%%%%%%%%%%%%%%%%%%%%%%%%%%
%%%%%%%%%%%%%%%%%%%%%%%%%%%%%
%%%%%%%%%%%%%%%%%%%%%%%%%%%%%

Above the scale $M$, the gauge group has 4 additional
${\bf 5} \oplus \bar{\bf 5}$ compared to the MSSM, and the SM gauge
couplings  remain perturbative up to the GUT scale.
In the range of scales where the $SU(3)_{S'}$ gauge coupling 
is at a strong fixed point, the beta functions of the SM gauge
couplings are reduced compared to the tree-level ones, making the 
couplings even more perturbative: each of the 3 strongly coupled flavors
counts as $\frac{5}{7}$ flavors in the RG equations
for the SM gauge couplings.

We must take $f \lsim 160\GeV$ in order to avoid a Landau
pole for $y_t$, which requires $g_S \gsim 2.5$ at the weak
scale (see Fig.~\ref{fig:flam_simp}).
This is perfectly natural in this model if the $SU(3)_{S'}$ coupling
is at its fixed point down to scales of order $100\TeV$.
We note that this puts the model in an interesting parameter regime
that favors an induced tadpole, and therefore a small value of the 
Higgs cubic coupling.

This model is only an example.
We can imagine other kinds of new physics that replaces the Landau
pole of this coupling, such as compositeness or extra dimensions.
Such a UV completion preserves unification as long as all new
fields are in complete $SU(5)$ multiplets and there are not too
many of them.
It is also possible to have larger values of $f$ if the top
quark is composite, as in the models of \Ref{Strassler:1995ia}.
These models generate complete composite GUT multiplets, and
therefore also preserve unification.

% =============================================================================
\subsection{Discussion
\label{sec:discussion}}
% =============================================================================
The essential ingredient in this model is the presence of additional
`auxiliary' Higgs fields charged under a new non-abelian gauge group $SU(2)_S$.
The new gauge coupling can easily be larger than the electroweak
gauge couplings at the TeV scale, so these Higgs fields naturally have
a large quartic coupling.
This model can be viewed as a perturbative (and hence calculable) version
of the mechanism of `superconformal technicolor' proposed in \Ref{Azatov:2011ps,Azatov:2011ht}.
In those models, the auxiliary Higgs fields are composites arising from strong
confining dynamics at the TeV scale, similar to technicolor.
The presence of strong dynamics at the same scale as the SUSY breaking
scale is not a coincidence if the strong dynamics is conformal above
the TeV scale, and confinement and electroweak symmetry breaking
are triggered by soft SUSY breaking.
Precision electroweak corrections are not precisely calculable in this
model, but using `\naive\ dimensional analysis' estimates place them near
the edge of the 95\% confidence level constraints.

The present perturbative models are fully calculable, so we can check all
experimental constraints without large theoretical uncertainties.
But it is not at all clear that nature cares about whether we can calculate.
One sense in which the present models are an improvement
over the strongly-coupled models is that they are compatible
with perturbative unification.
This comes at a price however, since putting the auxiliary Higgs fields
in a ${\bf 5} \oplus \bar{\bf 5}$ of $SU(5)$ means that their VEVs break custodial
symmetry.%
\footnote{The order parameter used in the
superconformal technicolor models in \Ref{Azatov:2011ps,Azatov:2011ht} was a $(2, 2)$ of
$SU(2)_L \times SU(2)_R$, which preserves custodial
symmetry but does not have a simple embedding into a complete GUT multiplet.}
This necessitates a little hierarchy between the $SU(2)_S$ breaking
scale ($\sim \text{TeV}$) and the Higgs mass scale ($\sim 100\GeV$).
If we did not require gauge coupling unification, we could make a
simpler model where the auxiliary Higgs fields are $SU(2)_W$ triplets,
but we will not pursue this here.

In our discussion, we have
been agnostic about the superpartner spectrum.
In order to have a natural model, this spectrum must have a 
relatively light stop and gluino, and this is compatible with
experimental limits only for special spectra, such as a somewhat 
compressed spectrum.
We do not address how this arises, since the phenomenological
issues essentially factorize: our model is compatible with any
`natural' spectrum that is not ruled out by the data.

% =============================================================================
% =============================================================================
\section{$F$-term Models
\label{sec:Fterms}}
% =============================================================================
% =============================================================================
Next we consider the possibility that the quartic coupling of $\Si$ is generated by 
$F$ terms via an NMSSM-like superpotential interaction
\begin{align}
\eql{Fterm}
\Delta W = \la_S S \Si_u \Si_d,
\end{align}
where $S$ is a singlet and
$\Si_{u,d}$ are Higgs doublets.  
If we require perturbativity up to a high scale such as the GUT scale, 
the coupling $\la_S$ can be somewhat larger than the corresponding coupling in
the NMSSM because the top Yukawa coupling does not contribute to the leading order
RG running of $\la_S$.
The largest value of the singlet coupling compatible with perturbativity up to
the GUT scale is approximately $\lambda_S = 0.92$.
We will see that this is not large enough to make a realistic model
of induced EWSB, so we will also discuss a hybrid model with both $D$ and $F$-term
contributions to the auxiliary Higgs quartic that satisfies all
phenomenological constraints and is perturbative up to the GUT scale.

The $F$-term model is interesting also because it is has different phenomenology
than the $D$-term models.
One important difference is that it
requires a nonvanishing VEV for both $\Si_u$ and $\Si_d$ in order
for the quartic \Eq{Fterm} to play a role in EWSB.
We consider the case where the VEVs are given by
\begin{align}
\eql{bidoubletvev}
\langle \Si_u \rangle = \frac{1}{\sqrt 2} \begin{pmatrix} 0 & 0 \cr  f_u & 0 \end{pmatrix} ; \quad
\langle \Si_d \rangle =  \frac{1}{\sqrt 2} \begin{pmatrix} 0 &  f_d  \cr 0 & 0 \end{pmatrix}  . 
\end{align}
We will make the simplifying assumption that $\avg{S} = 0$.
Imposing this exactly is unnatural,
since gaugino loops will generate a nonzero
$A$ term of the form $S \Si_u \Si_d$, and this will give a tadpole
for $S$.
However, $\avg{S}$ can be suppressed if the $S$ mass-squared is somewhat
larger than the other soft masses, so this can be a good approximation.
This can be motivated phenomenologically from the fact that a large VEV for 
$S$ correlates with large mixing between  $H$ and  $S$, which is
constrained by the measured Higgs couplings.
We stress however that we are making this assumption mainly for
simplicity.
We believe that allowing $\avg{S} \ne 0$ will not significantly change
the main conclusions below.%

With this assumption, the potentially relevant terms in the
superpotential are
\beq
\De W = \mu H_u H_d + \la_u H_u \Si_d \Phi 
+ \la_d H_d \Si_u \Phi
+ \bar{\la}_u H_u \Si_d \bar\Phi
+ \bar{\la}_d H_d \Si_u \bar\Phi.
\eeq
as well as soft SUSY breaking terms
\beq\eql{FtermSUSYbreakterms}
\begin{split}
\De V &= m_{H_u}^2 |H_u|^2 + m_{H_d}^2 |H_d|^2 
+ m_{\Si_u} |\Si_u|^2 + m_{\Si_d} |\Si_d|^2 
\\
&\qquad
{}+ B\mu H_u H_d + B_S \Si_u \Si_d
+ B_u H_u \Si_d + B_d H_d \Si_u + \hc
\\
&\qquad
{}+ \text{$A$ terms}.
\end{split}
\eeq
The fact that we need an explicit $\mu$ term is due to our
simplifying assumption that $\avg{S} = 0$.
We further neglect the couplings $\la_{u,d}$ and $\bar{\la}_{u,d}$
and the corresponding $A$ terms,
also for simplicity.
For completeness we quote the resulting minimization conditions:
\begin{align}
\!\!\!\!
m_{H_d}^2 &= B\mu \tan \be - B_d \frac{s_{\be_\Si}}{c_\be}\cot \ga - \sfrac{1}{2} m_Z^2 \left(s_\ga^2 \cdot \cos 2 \be + c_\ga^2 \cdot \cos 2 \be_\Si \right),
\\
m_{H_u}^2 &= B\mu \cot \be + B_u \frac{c_{\be_\Si}}{s_\be}\cot \ga + \sfrac{1}{2} m_Z^2 \left(s_\ga^2 \cdot \cos 2 \be + c_\ga^2 \cdot \cos 2 \be_\Si \right),
\\
m_{\Si_d}^2  &= B_s \tan {\be_\Si} + B_u \frac{s_\be}{c_{\be_\Si}} \tan \ga
- \sfrac{1}{2} m_Z^2 \left(s_\ga^2 \cdot \cos 2 \be + c_\ga^2 \cdot \cos 2 \be_\Si \right)
\nonumber\\
&\qquad\qquad
{}- \sfrac{1}{2} \la_S^2 v^2 c_{\be_\Si}^2 c_\ga^2,
\\
m_{\Si_u}^2  &= B_s \cot \be_\Si -B_d \frac{c_\be}{s_{\be_\Si}} \tan \ga
+ \sfrac{1}{2} m_Z^2 \left(s_\ga^2 \cdot \cos 2 \be + c_\ga^2 \cdot \cos 2 \be_\Si \right)
\nonumber\\
&\qquad\qquad
{}- \sfrac{1}{2} \la_S^2 v^2 c_{\be_\Si}^2 c_\ga^2,
\end{align}
where
\beq
\tan{\be_\Si} = \frac{f_u}{f_d}.
\eeq
As in \S\ref{sec:Dterms} we consider a simplified case where
$v_d \ll v_u, \sqrt{f_u^2 + f_d^2}$, ({\it i.e.}~large $\tan\be$).
We have a mass matrix in the basis $(H_u^0, \Si_u^0, \tilde \Si_d^0)$ given by (denoting $\la_S = \la$)
\begin{align}
\mathcal M^2 = \begin{pmatrix}
m_{H_u}^2 & 0 & -\frac{m_{H_u}^2 v_u}{f_d} \vphantom{\biggl|} \\
0 &  \frac 12 \la^2 f_d^2 + \frac{m_{\Si_d}^2 f_d^2}{f_u^2}  - \frac{m_{H_u}^2 v_u^2}{f_u^2} & \frac 12 \la^2 f_u f_d - \frac{m_{\Si_d}^2 f_d}{f_u} +  \frac{m_{H_u}^2 v_u^2}{f_u f_d}\\
 -\frac{m_{H_u}^2 v_u}{f_d}   &  \frac 12 \la^2 f_u f_d - \frac{m_{\Si_d}^2 f_d}{f_u} +  \frac{m_{H_u}^2 v_u^2}{f_u f_d}  & \frac 12 \la^2 f_u^2 + m_{\Si_d}^2 \vphantom{\biggl|}
 \end{pmatrix}.
\end{align}

\begin{figure}[hbt]
\begin{center}
\includegraphics[height=6cm]{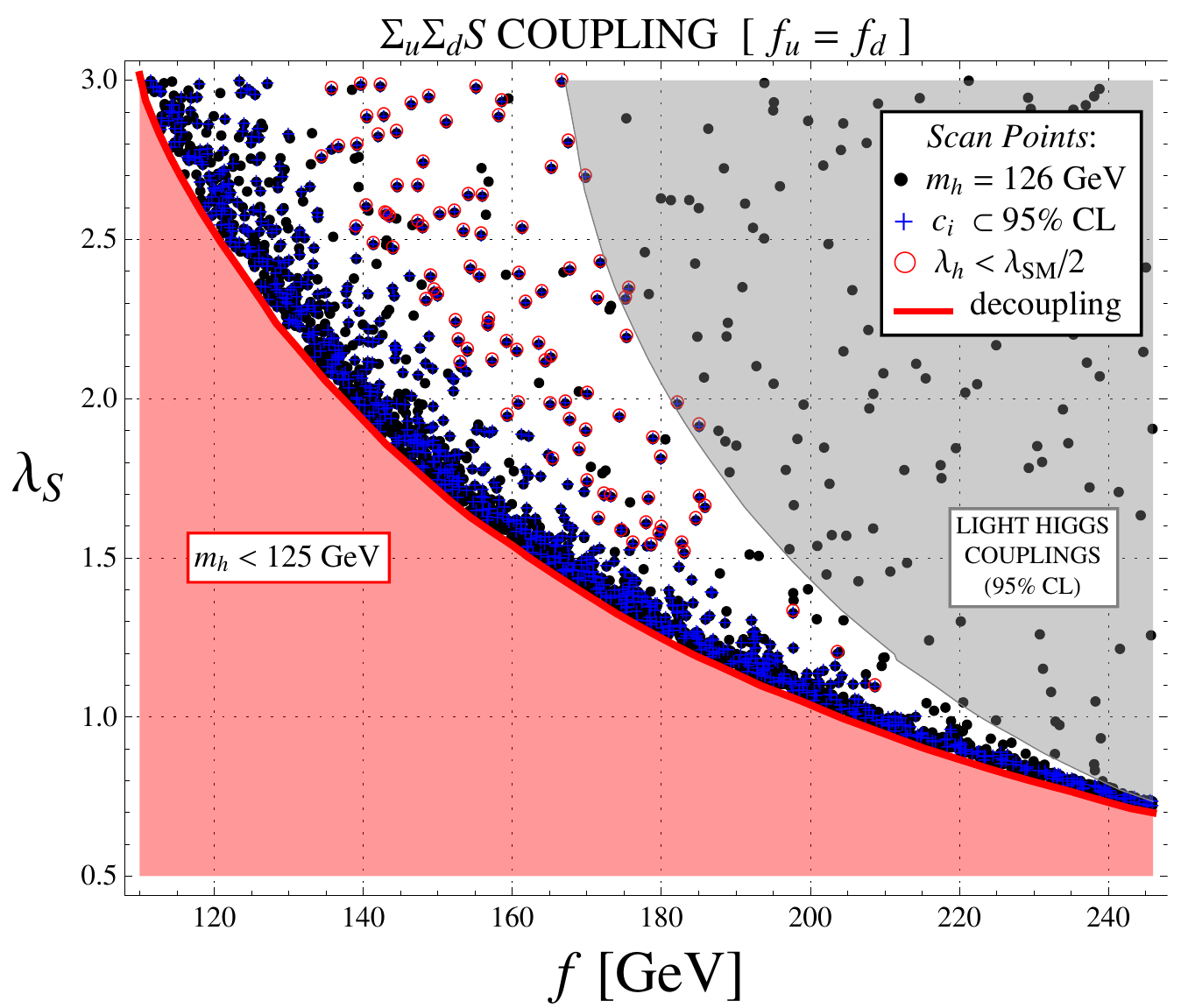} 
\includegraphics[height=6cm]{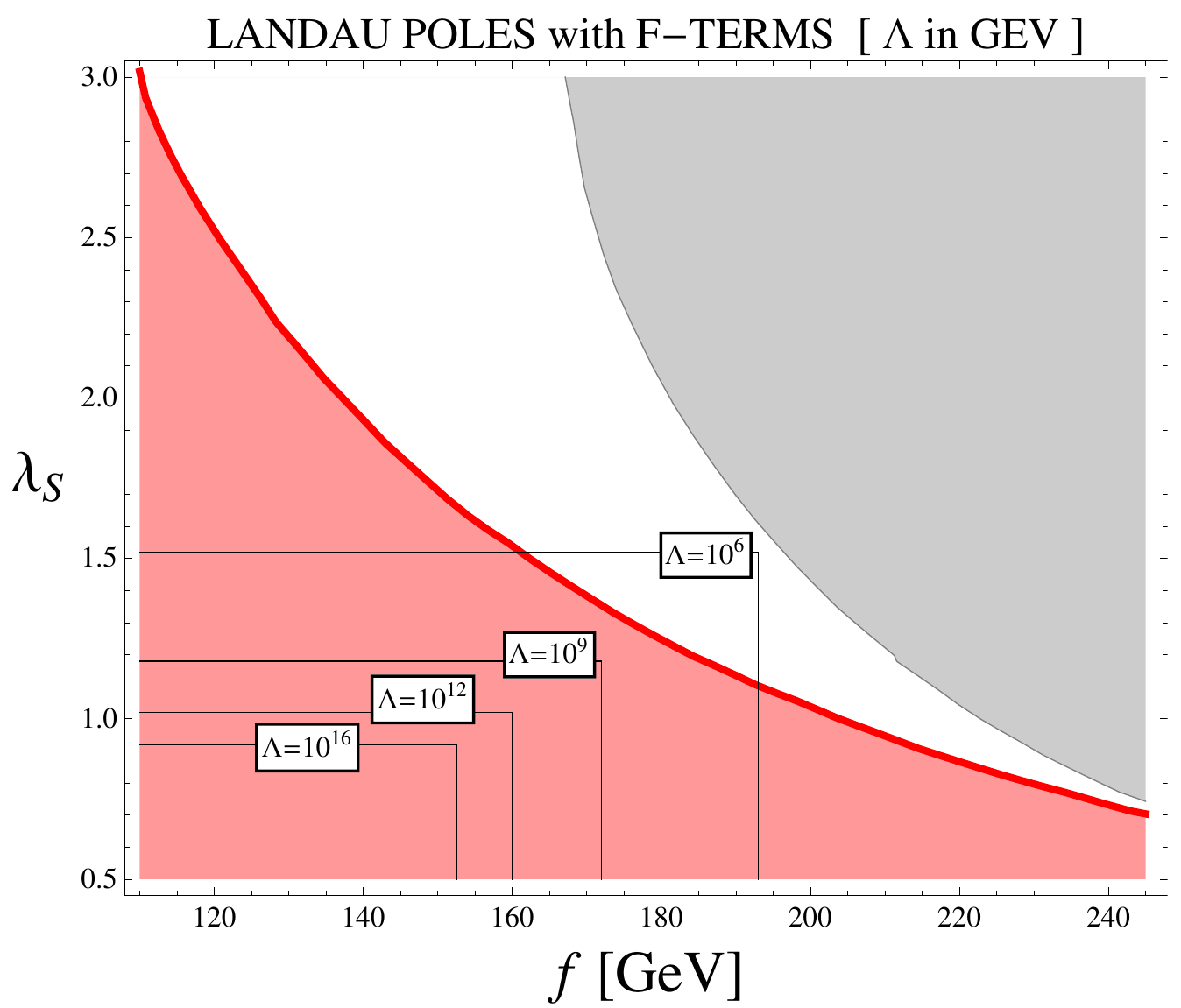} 
\caption{\small Parameter space of the model with non-decoupling $F$-terms.  {\bf Left:} Numerical scan of parameter space; the points are labeled as in Fig.~\ref{fig:Dterms_tb5}.  {\bf Right:} Position of Landau poles (compare Fig.~\ref{fig:DDrunning}).}
\label{fig:Fterms}
\end{center}
\end{figure}

This simplified model has 9 parameters, given by the mass terms in
\Eq{FtermSUSYbreakterms} and the superpotential coupling $\la_S$.
We trade 5 of these parameters for the VEVs $v_{u,d}$ and $f_{u,d}$
and the lightest Higgs mass and scan over the remaining parameters.
We display the results in the plane of $\la_S$ and $f = \sqrt{f_u^2 + f_d^2}$
in Fig.~\ref{fig:Fterms}.
As we expect, the results are qualitatively similar to both the
simplified model and the $D$-term model.
In particular, the Higgs VEV is dominated by an induced tadpole for roughly
half of the allowed parameter space.
The Landau poles generally occur at lower scales in this model than in the
$D$-term model.
One important difference is that $\la_S$ coupling runs at one loop, while 
the gauge coupling $g_S$ in the $D$-term model runs only at 2 loops.
For $y_t$ there are two effects. 
First, $y_t$ must typically be larger in the $F$-term models because
both $f_{u,d}$ are nonzero.
Second, there are 2 additional ${\bf 5} \oplus \bar{\bf 5}$ in the $D$-term model,
and only one in the $F$-term model.
This  makes the  $SU(3)_C$ gauge coupling larger at high scales
in the $D$-term model, slowing the running of $y_t$.
We could of course add additional ${\bf 5} \oplus \bar{\bf 5}$ to the $F$-term model,
but we will not explore this here.

\begin{figure}[hbt]
\begin{center}
\includegraphics[height=6cm]{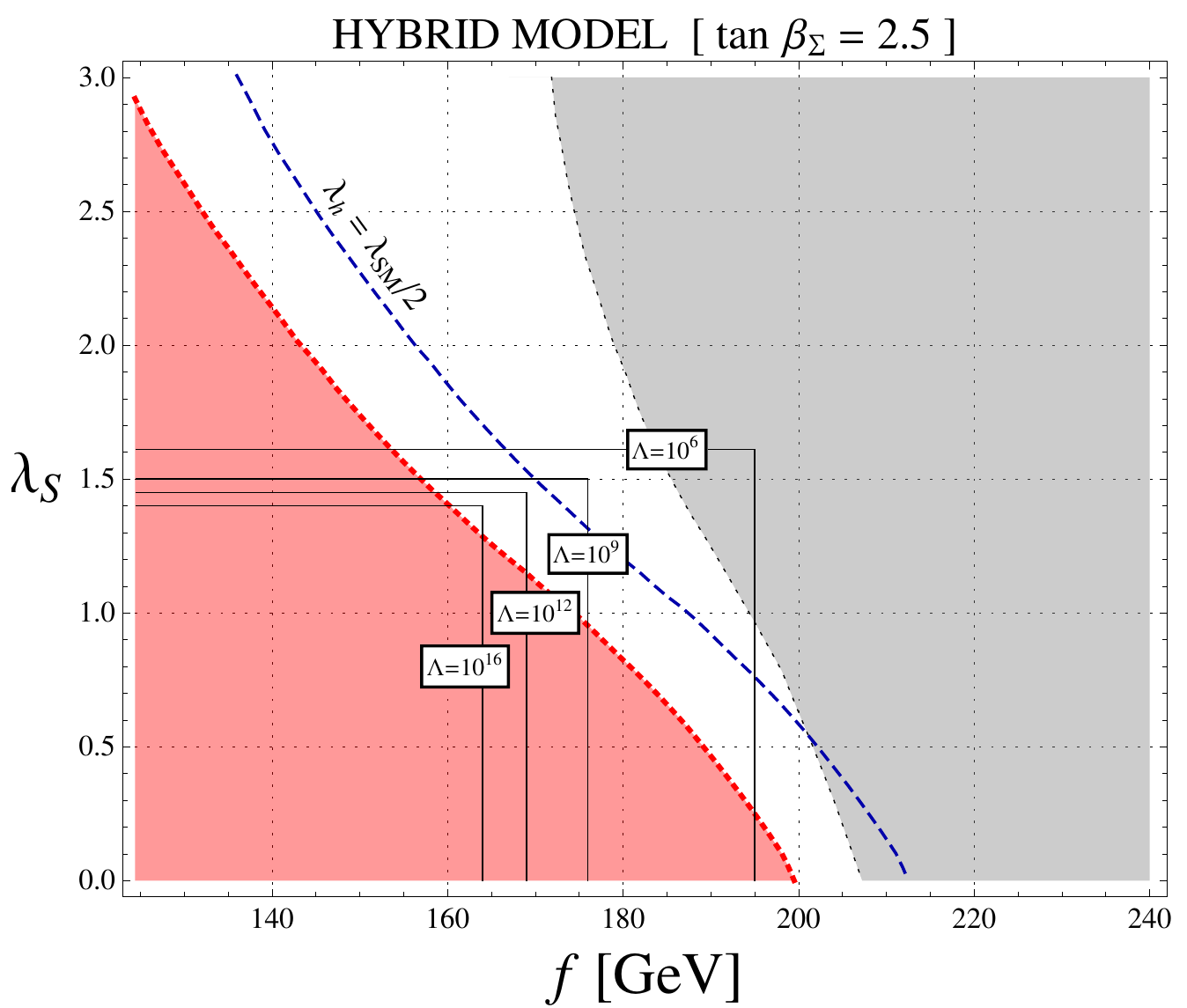} 
\caption{\small Allowed region and Landau poles (corresponding to a given point's cutoff) for a hybrid model with $F$ and $D$ terms.  The unshaded area corresponds to the allowed space, and cutoff values are quoted in GeV.  The position of the Landau poles is determined by optimizing the value of the couplings $g_S(\mu = m_W)$ and  $\lambda_\Phi(\mu =m_W)$ at each point.}
\label{fig:pert}
\end{center}
\end{figure}

To get a model compatible with perturbative unification, we turn to a
`hybrid' model where both $F$ and $D$ terms contribute to the auxiliary Higgs quartic.
%%%%%%%%%%%%%%%%%%%%%%%%%%%%%
%%%%%%%%%%%%%%%%%%%%%%%%%%%%%
%%%%%%%%%%%%%%%%%%%%%%%%%%%%%
%%%%%%%%%%%%%%%%%%%%%%%%%%%%%
In the model we consider, we choose an intermediate ratio of auxiliary VEVs, 
$\tan \be_\Si = 2.5$, which allows the singlet coupling $\la_S$ to contribute 
significantly to the auxiliary quartic.
The weak scale value of $\la_S$ can be larger than in the pure $F$-term model
because $g_S$ gives a negative contribution to the $\la_S$ beta function.
The resulting picture is shown in Fig.~\ref{fig:pert}.  
We see that we get a model that is perturbative up to the GUT scale only 
if the model is close to the decoupling limit.
This may be regarded as a kind of tuning, but see the discussion below.
Models with additional structure below the GUT scale may be a more attractive
possibility to combine naturalness with unification.
As in the $D$-term models, the auxiliary Higgs fields naturally come in complete
$SU(5)$ multiplets, so UV completing these models in a manner that 
preserves gauge coupling
unification should be possible, but we will not attempt to address that
here.

We conclude by commenting on the naturalness of the $F$-term models discussed above.
Models with $D$-terms have a tree-level violation of custodial symmetry
from the $\Si$ VEVs.
This is absent in the pure $F$-term model, and so this model requires no
little hierarchy among the masses, and therefore there is no danger
of  fine-tuning.
This is certainly a very attractive feature of this model.
In the hybrid model, we must have a little hierarchy to avoid 
large $T$-parameter corrections, but the value of $g_S$ is smaller,
so the tuning is reduced compared to the pure $D$-term model.

%%%%%%%%%%%%%%%%%%%%%%%%%%%%%
%%%%%%%%%%%%%%%%%%%%%%%%%%%%%
%%%%%%%%%%%%%%%%%%%%%%%%%%%%%
%%%%%%%%%%%%%%%%%%%%%%%%%%%%%

% =============================================================================
% =============================================================================
\section{Conclusions}
\label{sec:Conclusions}
% =============================================================================
% =============================================================================
We have presented SUSY models that address the Higgs naturalness problem.
In these models the dominant source of electroweak
symmetry breaking is due to the VEVs of the MSSM Higgs fields $H_{u,d}$,
but there is a subleading contribution to electroweak symmetry breaking
from additional `auxiliary' Higgs fields.
That is, we have
\beq
m_W^2 = \sfrac 14 g^2 (v_u^2 + v_d^2 + f^2),
\eeq
where $f$ arises from the auxiliary Higgs sector.
A typical value is 
$f = 150\GeV$, which gives $\sqrt{v_u^2 + v_d^2} = 195\GeV$.
The auxiliary Higgs fields have no Yukawa couplings, which allows them
to have a large quartic coupling either from additional gauge interactions
or superpotential interactions.
The masses of the auxiliary Higgs bosons are therefore above 125~GeV,
even though their contribution to electroweak symmetry breaking is subleading.
One simple limit of this model is the decoupling limit, where the
heavy mass eigenstate has a very small VEV.
In this limit, integrating out the auxiliary Higgs fields generates an
induced quartic for the MSSM Higgs fields that can explain the observed
125~GeV Higgs particle.
The light Higgs has standard model-like couplings in the decoupling limit,
so this limit is compatible with all the current Higgs data.

However, the data allows significant deviations from the decoupling limit.
In fact, we can access another limit where the heavy auxiliary Higgs
mass eigenstate has a significant VEV.
In this case, electroweak symmetry is nonlinearly realized in the
effective theory below the auxiliary Higgs mass.
This allows the effective theory to contain relevant electroweak breaking
terms such as tadpole terms for the light Higgs fields.
If this tadpole is sufficiently large, the light Higgs VEVs and masses are very
insensitive to the quartic coupling of the light Higgs.
This means that the quartic (and cubic) coupling of the Higgs can be much
smaller than those of the standard model Higgs.
This is a smoking-gun signal of this mechanism.

Between these two limits there is a large and phenomenologically
interesting parameter space.
Importantly, the tuning is less than $10\%$ in almost all of the
allowed parameter space, making this solution to the tuning problem
more robust than many others considered in the literature.

These models have rich phenomenology at the LHC and beyond.
First, the Higgs couplings can have significant deviation from their standard-model
values away from the decoupling limit.
Away from this limit, the auxiliary Higgs particles can be light,
and mix significantly with the light Higgs fields.
The auxiliary Higgs fields have suppressed couplings to fermions and
gauge bosons, but can have appreciable production cross sections.
They can decay to electroweak gauge bosons and/or the 125~GeV Higgs.
Current standard Higgs searches in the heavy mass region however do not yet 
have sufficient sensitivity to constrain the multi-Higgs models presented here: 
considering a benchmark mass of 350 GeV for the heavy state and assuming all final 
states are the same as those of the light Higgs, the current exclusions reach no lower 
than $\mu = \Ga / \Ga_\text{SM} \simeq 0.2$ 
combining all $H \to VV$ channels, corresponding to $H$ couplings 
of order $45\%$ or less of the corresponding couplings of the light 
Higgs\cite{Chatrchyan:2013yoa}.  
This is easily compatible with light Higgs couplings that are within 
$10\%$ of the SM, so both direct and indirect probes of the heavy states still allow 
for their (undetected) existence.  
Moreover, there can be a significant branching fraction for heavy Higgs to decay
to lighter Higgs states, such as $H \to hh$ if the heavy states are not decoupled
\cite{Craig:2013hca}, which further weaken direct search bounds.
The detailed bounds depend on the full parameter space of the model, not
just the 2-parameter subspace emphasized here, and goes beyond the scope of
this study.

Another important point for Higgs phenomenology is that the
light Higgs quartic (and hence cubic) coupling can 
easily be highly suppressed compared to the standard model.
Because of the destructive interference with direct double Higgs production,
this gives an increased rate for double Higgs production compared to the
standard model, which may be observable at the 14~TeV LHC
with 300~fb$^{-1}$ \cite{Klute:2012pu, Dolan:2012rv, Goertz:2013kp} and may 
provide additional evidence for the class of models studied here.

Finally, there is the rest of the superpartner spectrum.
The present model is motivated by naturalness, so the stops cannot 
be too heavy in this model.
Minimal predictive models of SUSY breaking (such as gauge mediation,
gaugino mediation, or anomaly mediation) do not give a natural
allowed SUSY spectrum, so we simply treat the soft SUSY breaking
terms as phenomenological parameters.
The superpartner spectrum is constrained by the absence of a signal
in SUSY searches so far.
There are a number of ways that this could happen, including 
(but not limited to)
$m_{\tilde{t}} \sim m_t$,
a compressed spectrum, 
$R$-parity violation,
or decays through a hidden sector \cite{Fan:2011yu}.
Searches at the 14~TeV LHC will have a large reach in these
scenarios, and we fervently  hope for a signal there.

% =============================================================================
% =============================================================================
\section*{Acknowledgements}
We have benefited from discussions with
A.~Azatov,
R.~Contino,
N.~Craig,
C.~Englert,
J.~Terning,
and N.~Weiner.
M.L. and Y.T. are supported by the Department of Energy under 
grant DE-FG02-91ER40674. 
J.G. is supported by the ERC Advanced Grant No.~267985.  Y.Z. is supported by SLAC, which is operated by Stanford University for the US Department of Energy under contract DE-AC02-76SF00515.
%{\it Electroweak Symmetry Breaking, Flavour and �Dark Matter: One Solution for Three Mysteries (DaMeSyFla)}.
% =============================================================================
% =============================================================================

% -----------------------------------------------------------------------------
\appendix{Appendix: Renormalization Group for Auxiliary Higgs Models}
% -----------------------------------------------------------------------------

Various order-1 couplings are required in each of the models presented.  
In practice we run couplings, including SM gauge couplings, at two loops with results that have been verified using the package {\tt SARAH} \cite{Staub:2009bi,Staub:2010jh}.
However, we summarize here the leading RG equations for three separate cases
to facilitate simple checks of the results:\ {\it i}) auxiliary quartics arising from $D$ terms alone; 
{\it ii}) quartics from $F$ terms alone; 
and {\it iii}) quartics from both $F$ and $D$ terms (i.e. the `hybrid' model). 
We have the following:
\begin{align}
D\mbox{-term model:} \quad  
\begin{cases} 
b_{g_S}^{(2)} \ \ =& 2 g_S^3 \left(g_1^2 + 3 g_2^2 + 8 g_3^2 + 9 g_S^2  -\la_\Phi ^2 \right),\\
b_{\la_\Phi}^{(1)} \ \ =& \la_\Phi \left( 4 \la_\Phi^2 - 3 g_S^2   \right), \\
%\be_{\la_\Phi}^{(2)} \ \ =& \la_\Phi \left(-10 \la_\Phi^4 + 6 g_S^2 \la_\Phi^2 + 9 g_S^4/2 \right) \\
\end{cases} \hspace{1.2cm}  \\
F\mbox{-term model:} \quad  
\begin{cases} 
b_{\la_S}^{(1)} \ \ =& \la_S \left( 4 \la_S^2 - \frac{3}{5} g_1^2  - 3 g_2^2  \right), \\
\end{cases} \hspace{3.1cm}  \\
\mbox{Hybrid model:} \quad 
\begin{cases} 
b_{g_S}^{(2)} \ \ =& 2 g_S^3 \left(g_1^2 + 3 g_2^2 + 8 g_3^2 + 9 g_S^2  -\la_\Phi ^2 - 2 \la_S^2 \right),   \\
b_{\la_\Phi}^{(1)} \ \ =& \la_\Phi \left( 4 \la_\Phi^2 - 3 g_S^2   \right), \\
b_{\la_S}^{(1)} \ \ =& \la_S \left( 6 \la_S^2  - \frac{3}{5} g_1^2  - 3 g_2^2 - 3 g_S^2   \right), \\
\end{cases}
\end{align}
with $\beta_g \equiv dg/d \ln \mu = \sum_\ell b_g^{(\ell)}/(16 \pi^2)^\ell$.
In the case of the extended $D$ term model where $SU(2)_S$ is embedded into the larger $SU(3)_{S'}$ at an intermediate scale $M$, we have  leading RG running in the (weakly-coupled) UV governed by
\bea
b_{\lambda_{\Phi_{1,2}}}^{(1)} &=& 
\lambda_{\Phi_{1,2}} \left( 5 \lambda_{\Phi_{1,2}}^2 + 3 \lambda_{\Phi_{2,1}} ^2- \frac{16}{3} g_{S'}^2\right), \\
b_{g_S}^{(1)} &=& -2 g_S^3 \, , \\
b_{g_S}^{(2)} &=&  g_{S'}^3\left(2 g_1^2 + 6 g_2^2 +16 g_3^2 + \frac{76}{3} g_{S'}^2 
-2 \left(\lambda_{\Phi_1}^2 + \lambda_{\Phi_2}^2 \right) \right) .\,
\eea
Here the superpotential 
couplings are given by
$W = \la_{\Phi_1} S \bar \Phi' \Phi' + \la_{\Phi_2} S \bar \Phi \Phi$.
When using these equations, it is important to take into account the
fact that there are extra multiplets on the running of the standard model
gauge couplings, which also affects the running of the top quark Yukawa coupling.

% Bibliography
%\bibliographystyle{utphys}
%\bibliography{Induced.bbl}

%\newpage

\end{document}